\documentclass[fleqn,useAMS,usenatbib]{mnras}

\usepackage[T1]{fontenc}
\usepackage{ae,aecompl}

\usepackage{pdflscape}
\usepackage{graphicx}	
\usepackage{amsmath}	
\usepackage{amssymb}	
\usepackage{color}

\usepackage{newtxtext,newtxmath}

\newcommand{\kms}{\hbox{${\rm km\;s}^{-1}$}}

\newcommand{\hi}{H~\textsc{i}}   
\newcommand{\ha}{H$\alpha$}

\newcommand{\amax}{\ensuremath{a_{\epsilon}}}
\newcommand{\amin}{\ensuremath{a_{\mathrm{min}}}}
\newcommand{\aten}{\ensuremath{a_{10}}}
\newcommand{\lbar}{\ensuremath{L_{\mathrm{bar}}}}

\newcommand{\pabox}{\ensuremath{\mathrm{PA}_{\mathrm{box}}}}
\newcommand{\rbox}{\ensuremath{R_{\mathrm{box}}}}
\newcommand{\deltapabar}{\ensuremath{\Delta \mathrm{PA}_{\mathrm{bar}}}}

\newcommand{\Msun}{\ensuremath{M_{\sun}}}
\newcommand{\Mstar}{\ensuremath{M_{\star}}}
\newcommand{\logmstar}{\ensuremath{\log \, (M_{\star}/M_{\sun})}}
\newcommand{\MHI}{\ensuremath{M_{\mathrm{H} \textsc{i}}}}
\newcommand{\Matomic}{\ensuremath{M_{\mathrm{H} \textsc{i} + \mathrm{He}}}}
\newcommand{\mhi}{\ensuremath{m_{21,\mathrm{c}}}}
\newcommand{\fgas}{\ensuremath{f_{\mathrm{gas}}}}
\newcommand{\bmv}{\ensuremath{B\! - \!V}}
\newcommand{\bmvre}{\ensuremath{(B\! - \!V)_{e}}}
\newcommand{\bmk}{\ensuremath{B\! - \!K}}
\newcommand{\gmr}{\ensuremath{g\! - \!r}}
\newcommand{\ho}{\ensuremath{H_{0}}}
\newcommand{\sfourg}{S\ensuremath{^{4}}G}

\newcommand{\fBPbars}{\ensuremath{f_{\rm bars}(\mathrm{B/P})}}

\title[Frequency of Boxy/Peanut-Shaped Bulges in Barred Galaxies]{The Frequency and
Stellar-Mass Dependence of Boxy/Peanut-Shaped Bulges in Barred Galaxies}
\author[P. Erwin \& V.P. Debattista]{Peter Erwin$^{1,2}$\thanks{E-mail: 
erwin@mpe.mpg.de} and Victor P. Debattista$^{3}$ \\
$^{1}$Max-Planck-Insitut f\"{u}r extraterrestrische Physik, Giessenbachstrasse, 85748 Garching, Germany \\
$^{2}$Universit\"{a}ts-Sternwarte M\"{u}nchen, Scheinerstrasse 1, D-81679 M\"{u}nchen, Germany \\
$^{3}$Jeremiah Horrocks Institute, University of Central Lancashire, Preston PR1 2HE, UK}

\date{Accepted XXX. Received YYY; in original form ZZZ}

\pubyear{2017}

\begin{document}
\label{firstpage}
\pagerange{\pageref{firstpage}--\pageref{lastpage}}
\maketitle

\begin{abstract}

From a sample of 84 local barred, moderately inclined disc galaxies, we
determine the fraction which host boxy or peanut-shaped (B/P) bulges
(the vertically thickened inner parts of bars). We find that the
frequency of B/P bulges in barred galaxies is a very strong function of
stellar mass: 79\% of the bars in galaxies with $\logmstar \ga
10.4$ have B/P bulges, while only 12\% of those in lower-mass
galaxies do. (We find a similar dependence in data published by
\citealt{yoshino15} for edge-on galaxies.) There are also strong trends
with other galaxy parameters -- e.g., Hubble type: 77\% of S0--Sbc bars,
but only 15\% of Sc--Sd bars, have B/P bulges -- but these appear to be
side effects of the correlations of these parameters with stellar mass.
In particular, despite indications from models that a high gas content
can suppress bar buckling, we find no evidence that the (atomic) gas
mass ratio $\Matomic/\Mstar$ affects the presence of B/P bulges, once
the stellar-mass dependence is controlled for.

The semi-major axes of B/P bulges range from one-quarter to
three-quarters of the full bar size, with a mean of $\rbox/\lbar = 0.42
\pm 0.09$ and $\rbox/\amax = 0.53 \pm 0.12$ (where \rbox{} is the size
of the B/P bulge and \amax{} and \lbar{} are lower and upper limits on the
size of the bar).

\end{abstract}


\section{Introduction}\label{sec:intro} 

Over the past three decades, various lines of evidence have converged to
demonstrate that many bulges in edge-on galaxies are at least partly
the vertically thickened inner parts of galactic bars (e.g.,
\citealt{laurikainen16} and \citealt{athanassoula16}, and references
therein). This phenomenon appears in the form of bulges with ``boxy'',
``peanut-shaped'', or ``X-shaped'' morphologies; we will refer to these
generically as boxy/peanut-shaped (B/P) bulges. Most if not all of the
Milky Way's own bulge is now understood to be just such a structure
\citep[e.g.,][and references
therein]{nataf10,shen10,saito11,ness12,wegg13,wegg15,zoccali16,
shen16,debattista17}.

The formation of bars in $N$-body simulations of galaxy discs is often
followed, usually within a Gyr or so, by a violent vertical buckling
instability which then settles down into a B/P bulge
\citep[e.g.,][]{raha91,merritt94,
debattista04,martinez-valpuesta04,martinez-valpuesta06,debattista06,
saha13}. Direct detection of ongoing buckling in two nearby barred
spirals was recently made by \citet{erwin-debattista16}. Alternatively,
bars may experience slower and more symmetric vertical thickening, via
the trapping of disc stars at vertical resonances, again leading to a
B/P bulge
\citep[e.g.,][]{combes81,combes90,quillen02,debattista06,quillen14}. In
either case, the result is a bar with a vertically thin outer structure
and a vertically thick inner structure; this inner structure is
supported by a variety of off-plane orbits, mostly thought to derive
from the planar $x_{1}$ orbits which support bars, though other orbit
families may also contribute
\citep[e.g.,][]{pfenniger84,pfenniger85,pfenniger91,patsis02b,portail15,
valluri16,abbott16}. When seen close to end-on (down the long axis of
the bar), this structure appears round and is not easily distinguishable
from a classical (spheroidal) bulge, but when seen side-on, it appears
as a boxy or even peanut-shaped stellar structure; in particularly
strong cases the latter can have an X-shape
\citep[e.g.,][]{combes90,pfenniger91,lutticke00b,athanassoula02}. At
intermediate orientations, the B/P bulge can still form a boxy shape in
projection, which is the reason for the shape of our Galaxy's bulge.

Attempts have been made to connect populations of B/P bulges to
populations of bars using edge-on galaxies
\citep{jarvis86,de-souza87,shaw87,dettmar90,lutticke00a,yoshino15}.
However, these attempts suffer from the key problem that it is very
difficult to determine whether an edge-on galaxy has a (planar) bar. The
presence of a B/P bulge can (in most cases, at least) be taken as
indicating the presence of a bar; but the \textit{absence} of a visible
B/P bulge may be due to the orientation of the bar, or to a bar lacking
a B/P bulge, or to the absence of a bar altogether. Clearly, it would be
advantageous to be able to reliably identify B/P bulges in
\textit{non}-edge-on galaxies, where the separate question of whether a
planar bar is present can much more easily be answered.

Although the link between bars and B/P bulges was first made with
edge-on galaxies, recent studies have shown that B/P bulges can be
detected in moderately inclined and even face-on galaxies. Ways of doing
this include stellar-kinematic signatures in face-on galaxies
\citep{debattista05,mendez-abreu08} and specific bar morphologies in
galaxies with intermediate inclinations ($i \sim 40$--80\degr). The
advantage of face-on or moderately inclined galaxies is that the B/P
bulge can be related to the structure of the bar as a whole, as well as
to other structures in the disc.

Early evidence for morphological signatures of B/P bulges in galaxies with
inclinations of $\sim 70$--80\degr{} was presented by \citet{bettoni94}
for NGC 4442 ($i = 72\degr$) and \citet{quillen97} for NGC~7582 ($i =
68\degr$). \citet{athanassoula06} compared near-IR images of M31 ($i =
77\degr$) with projected $N$-body simulations to argue that much of the
(boxy) bulge in that galaxy was due to a bar.

By analyzing $N$-body simulations that formed B/P-bulge-hosting bars,
\citet{erwin-debattista13} showed that the projections of bars with B/P
bulges created a characteristic morphology in the isophotes of galaxies
with $i \sim 40$--70\degr. This morphology consists of a thick, often
``box''-shaped region, due to the B/P bulge itself, and thinner, offset
``spurs'' at larger radii due to the outer (vertically thin) part of the
bar (as had already been argued by \citealt{athanassoula06} for the
case of M31). We also showed how it was possible to relate measurements
of the box-shaped region to the linear extent of the B/P bulge along the
bar's major axis, making it possible to estimate the fraction of the bar
length which was taken up by the B/P bulge. Finally, we presented two
examples of galaxies with both strong bars and ideal orientations
for detecting B/P bulges which did \textit{not} show the signatures of
B/P bulges, but instead strongly resembled simulations with flat,
\textit{unbuckled} bars. This last result shows how it is possible to
identify bars \textit{without} B/P bulges.

Detection of B/P bulges in \textit{face-on} galaxies (e.g., $i
\lesssim 30\degr$) is also feasible using stellar-kinematic signatures
\citep{debattista05,mendez-abreu08,iannuzzi15}, though this does require expensive
spectroscopic observations. \citet{laurikainen14} and
\citet{athanassoula15} have shown that it is also possible to identify
B/P bulges in face-on galaxies using the ``barlens'' morphological
feature \citep{laurikainen11}.

It is thus now possible to survey galaxies with low or moderate
inclinations to determine whether they have B/P bulges or not, to
measure the sizes of B/P bulges, and to relate B/P bulges to their
parent bars and host galaxies. This makes it possible to address a number
of questions: How important are B/P bulges in the larger scheme
of galaxy morphology? What fraction of bars produce B/P bulges? (Do
\textit{all} bars have B/P bulges?) If some bars have them and some do
not, what if anything about the galaxy determines this? Is it unusual
for a galaxy like our own to have a B/P bulge, or is our galaxy typical
in this respect? In addition, by having an unbiased set of measurements
of B/P bulge sizes relative to their parent bars, we can potentially
place constraints on both their structure -- which sets of 3D orbits
actually support B/P bulges -- and their formation mechanisms.

\section{Sample Definition}\label{sec:sample} 

Our sample selection is dictated partly on our preferred approach
for identifying and measuring B/P bulges (see
Appendix~\ref{app:methods}), which requires galaxies with moderate
inclinations: neither edge-on nor too close to face-on. We
defined an initial sample by selecting all galaxies from the RC3 catalog
\citep{rc3} which met the following criteria: Galactic latitude $|b| >
20\degr$, $V_{\rm GSR} \le 2000$ km~s$^{-1}$, Hubble types S0--Sd ($-3.4
\le T \le 7.4$), diameters $D_{25} \ge 3.0\arcmin$, and estimated
inclinations (based initially on 25th-magnitude axis ratios) of
40\degr--70\degr. We also included galaxies with Virgo Cluster Catalog
\citep{binggeli85} entries with $V_{\rm GSR} > 2000$ km~s$^{-1}$, since
the high velocity dispersion of the Virgo Cluster makes a redshift
cutoff less meaningful. The purpose of these criteria was to select
galaxies with large, well resolved bars which had favourable
orientations for identifying B/P bulges \citep{erwin-debattista13}. We
excluded a total of 53 galaxies for a variety of reasons: having
inclinations clearly outside the required range, despite their RC3 axis
ratios (based on, e.g., ellipse fits to publicly available images,
velocity-field studies from the literature, etc.); being strongly
distorted due to ongoing interactions or mergers; misclassifications;
absence of useful imaging or photometric data; etc. (See
Appendix~\ref{app:rejected}.) The final parent sample (``Parent'')
included 186 galaxies; these are listed in Table~\ref{tab:parent}.

These 186 galaxies were visually inspected, using near-IR images
whenever possible, for the presence of bars. We identified 118 galaxies
with bars, for a total barred fraction of $63.4^{+3.5}_{-3.6}$\%
(binomial 68\% confidence intervals); this constitutes our ``Barred''
subsample. (Three galaxies with small ``nuclear bars'' but no
large-scale bars -- NGC~1201, NGC~1553, and NGC~5194 -- were rejected;
see Appendix~\ref{app:nuclear-bars}.)

Finally, we identified a subset of the barred galaxies where the bar
orientation (measured using a combination of ellipse fits and visual
inspection) was most favourable for identifying projected B/P structures.
Specifically, this meant galaxies where the in-plane (deprojected)
position angle of the bar was $\le 60\degr$ away from the galaxy major
axis ($\deltapabar \le 60\degr$). This is critical for maximizing our
ability to detect B/P bulges, since projected B/P bulges are difficult
to identify when the bar is oriented near the minor axis
\citep{erwin-debattista13}. This produced a final
``good-position-angle'' (GoodPA) sample of 84 galaxies, which was the
basis for our search for B/P bulges.

\subsection{Sample Characteristics} 

In Figures~\ref{fig:sample-distances}--\ref{fig:sample-mstar} we show
distributions of sample properties for the Parent sample and the two
subsamples (Barred and GoodPA). Most of the galaxies are found at
distances $< 25$ Mpc (median distance = 16.9 Mpc). The sample is
weighted towards relatively massive galaxies (median $\logmstar =
10.49 \Msun$).

The figures indicate that the two subsamples are not
meaningfully biased relative to their parent samples: that is, the
barred galaxies do not differ significantly from the parent sample, and
neither do the barred galaxies with favourable bar position angles.
Kolmogorov-Smirnov and Anderson-Darling two-sample tests find no
evidence for differences in stellar mass between the barred and unbarred
galaxies, or between the barred galaxies with good bar position angles
and the other barred galaxies for either stellar mass or distance ($P$
values ranging from 0.11 to 0.95).

\begin{figure*}
\begin{center}
\includegraphics[scale=0.78]{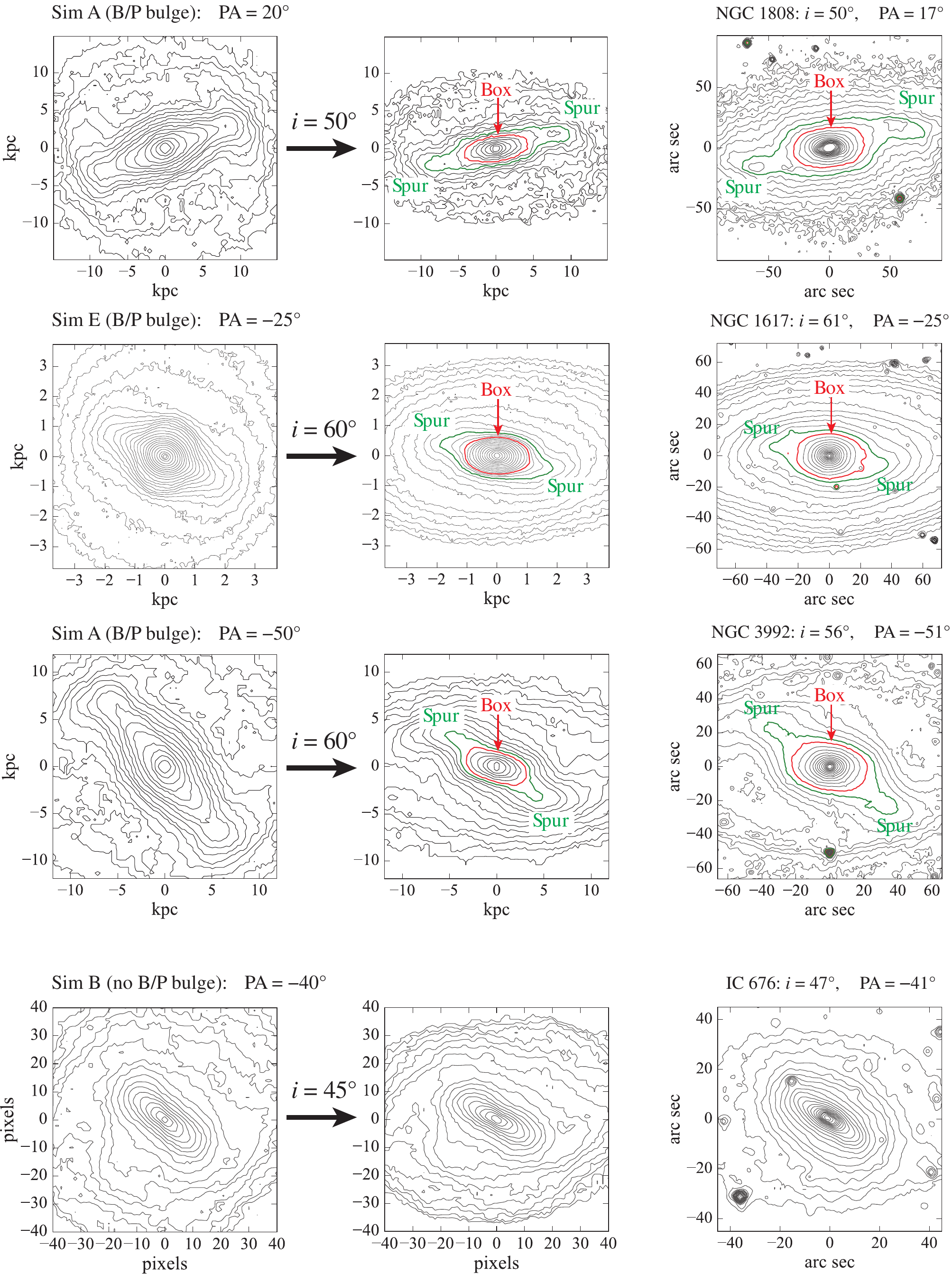}
\end{center}

\caption{Projection effects on isophotes for bars with and without B/P
bulges. Upper three rows: Left-hand panels show face-on views of
barred-galaxy simulations with B/P bulges, with bars rotated at
specified angles \deltapabar{} with respect to line of nodes
(horizontal); middle panels show same simulation projected at indicated
inclinations. Right-hand panels show real galaxies (\sfourg{}
3.6\micron{} images, except $H$-band from \citealt{eskridge02} for
NGC~1808) with similar orientations (rotated so disc major axes are
horizontal). Approximate regions of the ``box'' (boxy or oval projection
of the B/P bulge) are outlined in red and the ``spurs'' (projection of
the outer, vertically thin part of the bar) are outlined in green.  (See
\citealt{erwin-debattista13} for more examples.) Bottom row: Left:
Face-on view of simulation with \textit{vertically thin bar} (no B/P
bulge). Middle: same simulation at $i = 45\degr$. Right: SB0 galaxy
IC~676 (\sfourg{} image). Note that these bars show symmetric,
elliptical isophotes, with no box+spurs morphology, due to the
\textit{absence} of a B/P bulge. (See Appendix~\ref{sec:sims} for
details of the simulations.) \label{fig:box-spurs-demo}}

\end{figure*}

\begin{figure}
\begin{center}
\hspace*{-1.5mm}\includegraphics[scale=0.43]{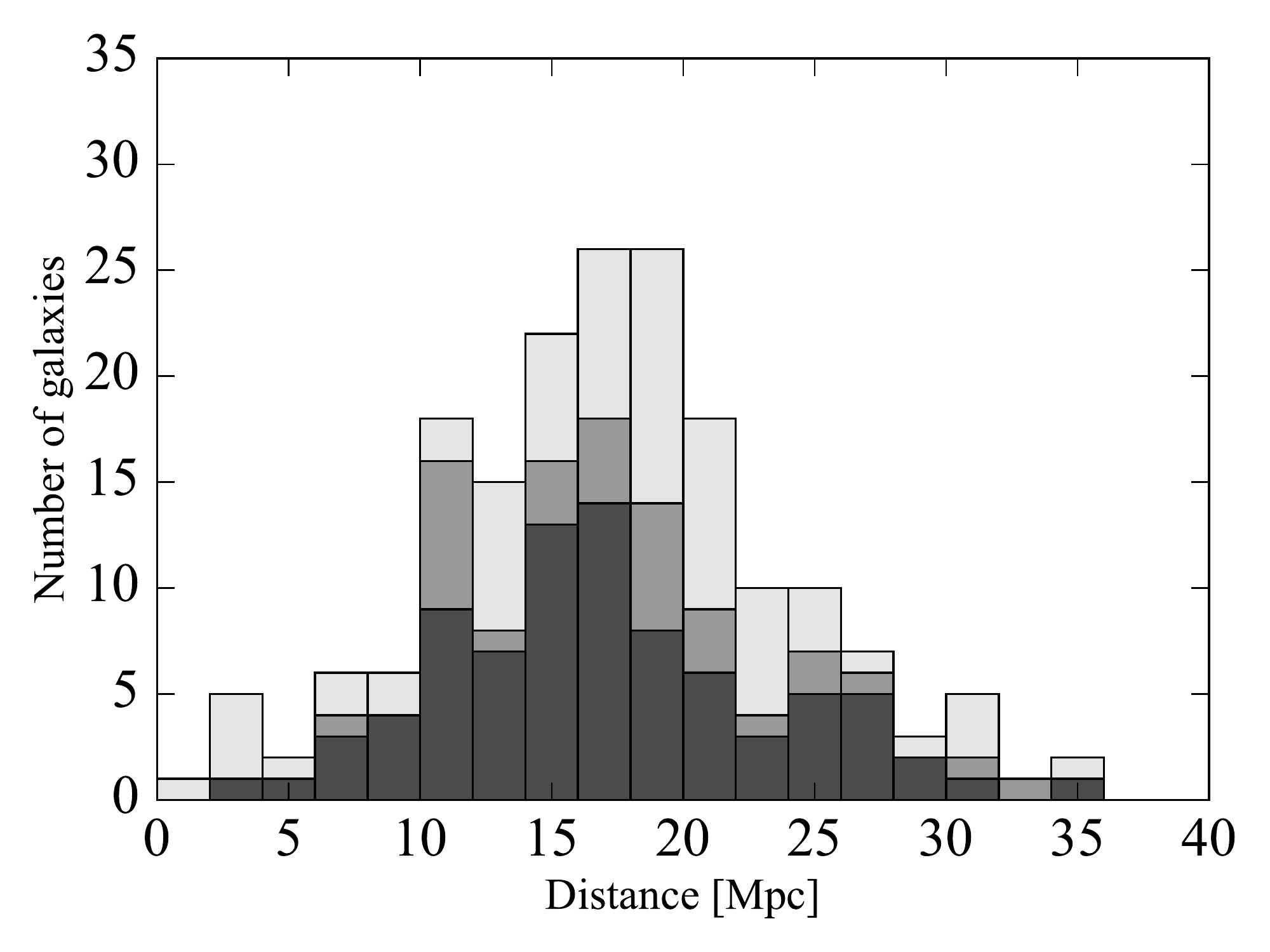}
\end{center}

\caption{Distribution of galaxy distances for the Parent sample (light grey)
and the Barred (medium grey) and GoodPA (dark grey) subsamples. \label{fig:sample-distances}}

\end{figure}

\begin{figure}
\begin{center}
\hspace*{-1.5mm}\includegraphics[scale=0.43]{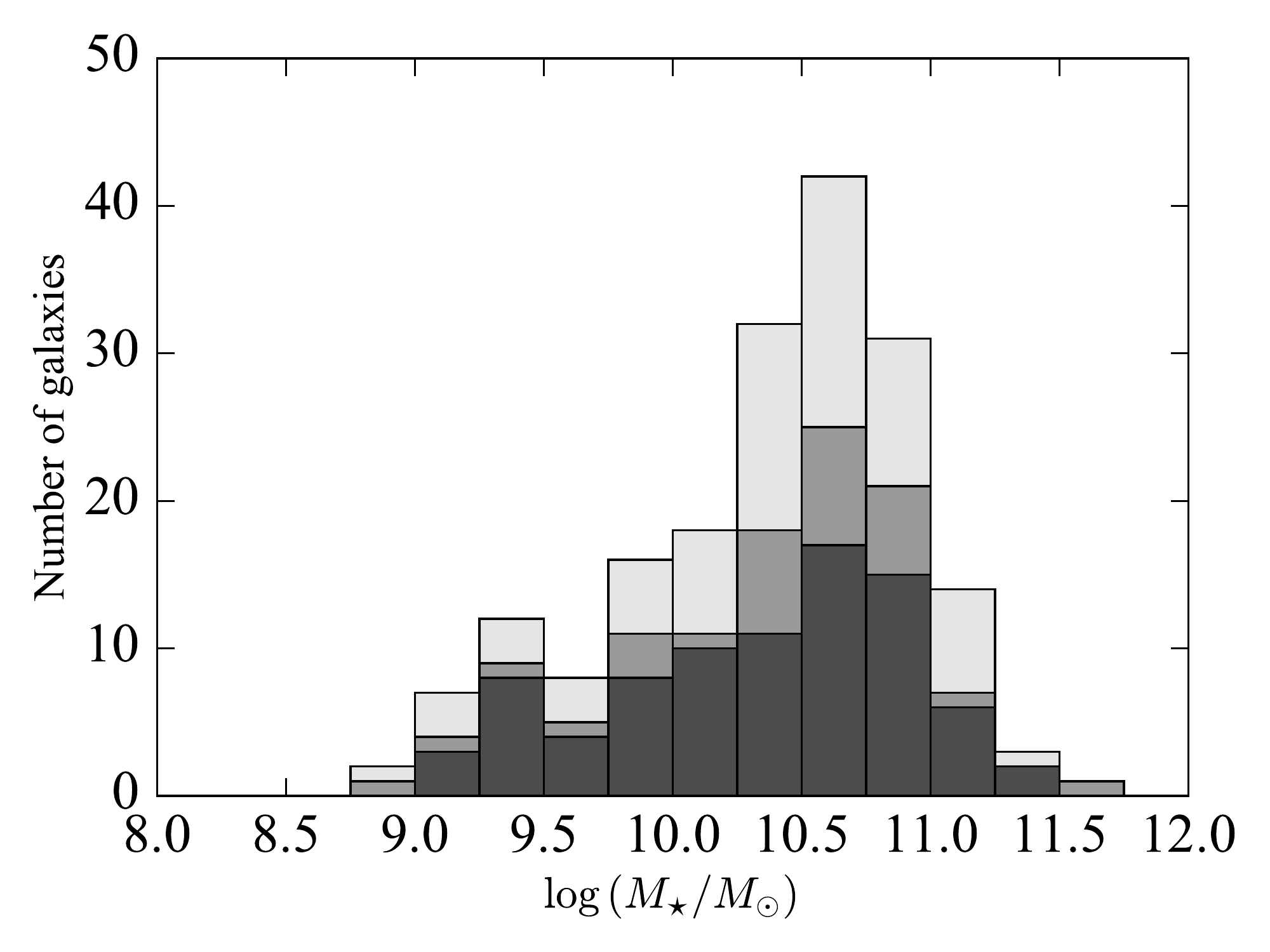}
\end{center}

\caption{As for Figure~\ref{fig:sample-distances}, but showing the
distribution of galaxy stellar masses.
\label{fig:sample-mstar}}

\end{figure}

\subsection{Data Sources}\label{sec:data} 

The data used for this project are of two types: optical or
near-IR images, which we use for detecting and measuring bars and B/P
bulges, and general physical parameters for the galaxies -- distances,
stellar masses, \bmv{} and \bmk{} colors, gas mass fractions ($\fgas =
\MHI / \Mstar$), and bar sizes -- which we use for investigating how the
presence or absence of B/P bulges might depend on galaxy properties.
Further details on the data sources and derived quantities are given in
Appendix~\ref{app:data}.

\section{The Identification and Measurement of B/P Bulges} 

Our basic approach for identifying B/P bulges inside bars is that
originally outlined in \citet{erwin-debattista13} and in the last part
of Appendix~\ref{app:methods}. This approach is based on the fact that a
bar which is vertically thin in its outer regions and vertically thick
in its interior produces a characteristic pattern in the isophotes when
viewed at intermediate inclinations: the B/P bulge projects to form
relatively thick, oval or box-shaped isophotes (the ``box''), while the
outer part of the bar projects to form thinner, offset isophotes (the
``spurs''). Because of this offset, the position angle of the spurs is
always further away from the galaxy major axis than that of the box/oval.
When the bar does \textit{not} have a vertically thick interior, it
projects instead to aligned, elliptical isophotes without the box + spurs
morphology.

Figure~\ref{fig:box-spurs-demo} presents examples of the box + spurs
morpholgy in the upper three rows and examples of bars without B/P bulges
in the bottom row; more examples can be found in
\citet{erwin-debattista13}. For most of our analysis, we classify the
two galaxies with \textit{currently buckling} bars \citep[NGC~3227 and
NGC~4569;][]{erwin-debattista16} as \textit{not} having B/P bulges,
though we identify them as a separate category in some of our figures.

\begin{figure*}
\begin{center}
\includegraphics[scale=0.85]{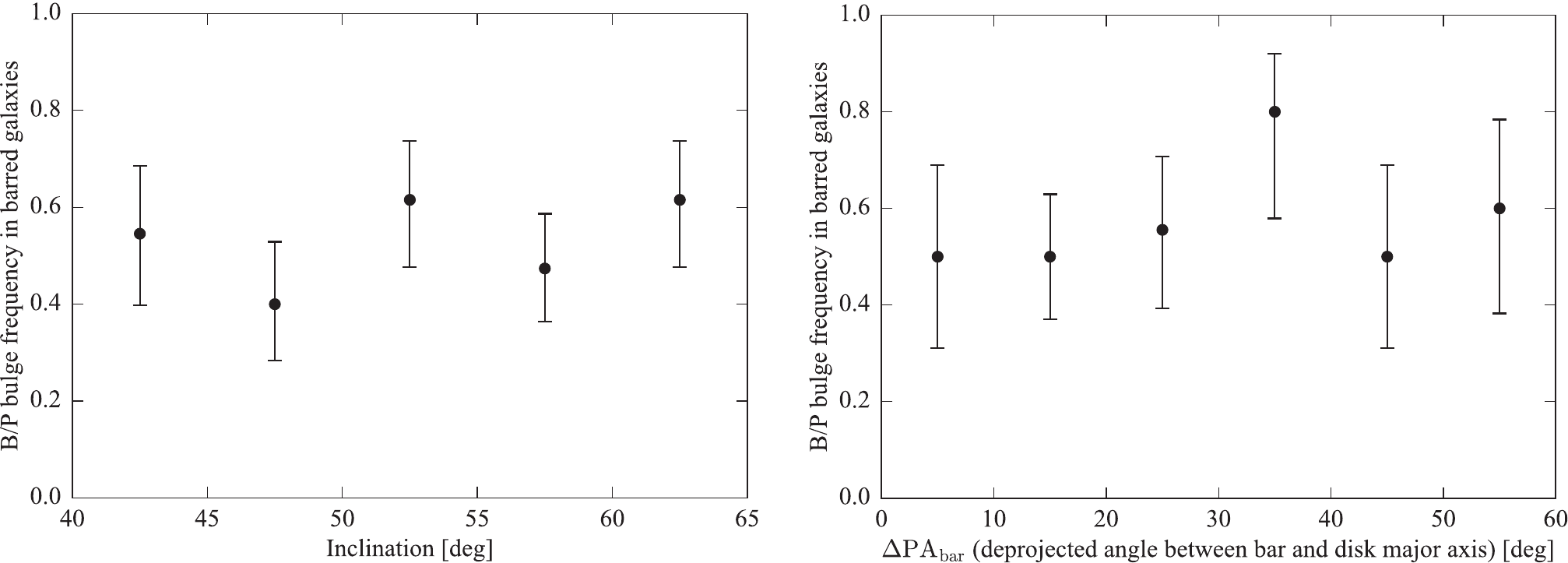}
\end{center}

\caption{Fraction of barred galaxies with detected B/P bulges, as a
function of galaxy inclination (left) and \deltapabar{} (deprojected
angle between bar and disc major axis, right). The absence of visible
trends suggests that the inclination and \deltapabar{} limits we use
for our analysis are
reasonable.\label{fig:fBP-vs-incPA}}

\end{figure*}

\subsection{Efficiency and Completeness Estimates (Are We Missing Many B/P Bulges?)} 

As we showed in \citet{erwin-debattista13}, the detectability of B/P
bulges depends strongly on the inclination of a galaxy. Galaxies too
close to face-on lack the projected difference between the vertically
thick B/P bulge and the vertically thin outer bar. When galaxies are too
highly inclined, it becomes difficult to discern the bar as a whole
(particularly the outer part) and to distinguish it from projected
rings, spiral arms, etc.

For galaxies with favourable inclinations, the position angle of the bar
with respect to the line of nodes -- that is, the observed major axis of
the galaxy disc -- is also important. Figures~5 and 6 of
\citet{erwin-debattista13} show that as the position angle of the bar
approaches the galaxy minor axis, the projection of the B/P bulge and
its contrast with the projected outer bar (the spurs) becomes steadily weaker and
more ambiguous. An additional problem is that extra axisymmetric
structures inside the bar (classical bulges, disky pseudobulges/nuclear
discs, nuclear rings) may overlap with the projected B/P bulge when the
bar is close to the minor axis \citep[see, e.g., Figs.~15--17
in][]{cole14}.  A lesser ambiguity may arise if the bar is very close to
($< 5\degr$ from) the \textit{major} axis, because then the spurs are no
longer visibly \textit{offset} from the axis of the inner bar or the
disc major axis.

In this study, we adopted an inclination range of 40--70\degr{} and
a relative bar-disc position-angle limit of $\deltapabar \le 60\degr$.
Was this too generous? In other words: is there evidence that we are
missing B/P bulges at inclinations near the lower or upper ends of the
inclination range, or at particularly large (or small) values of
\deltapabar{} bar? To check this possibility, we plot in the left panel
of Figure~\ref{fig:fBP-vs-incPA} the fraction of barred galaxies with
detected B/P bulges as a function of galaxy inclination. There are no
apparent trends in this figure, which suggests that we are \textit{not}
missing significant numbers of B/P bulges at either end of our
inclination range. The right panel of Figure~\ref{fig:fBP-vs-incPA}
indicates that our $\deltapabar$ limits are probably not biasing our
results, either -- in particular, our upper limit of $\deltapabar =
60\degr$ does not seem to have been too high.

\subsection{Measuring B/P Bulge Sizes} 

To estimate the sizes of B/P bulges, we follow the same approach used in
\citet{erwin-debattista13}: measuring the linear extent of the box/oval
along its major axis (\rbox), along with its position angle (\pabox). In
that paper, we showed that such measurements did a reasonably good job
of capturing the radial extent of the B/P bulge, as measured from
edge-on projections of $N$-body simulations (see Section~3.2 and Fig.~8
of \citealt{erwin-debattista13}).

Bar and B/P measurements are collected for all galaxies in the GoodPA
subsample in Table~\ref{tab:bars}. Visual indications of both \rbox{}
and \pabox{} for galaxies with new B/P detections (and galaxies
previously identified but not previously measured) are shown in
Figure~\ref{fig:new-bp-plots}. In Section~\ref{sec:size}, we compare our
measurements with other measurements of B/P bulge sizes from the
literature for the same galaxies and find generally good agreement.

\begin{table*}
\begin{minipage}{126mm}
    \caption{Bar and B/P Bulge Measurements}
    \label{tab:bars}
    \begin{tabular}{lrrrrrrrrrrr}
\hline
Name       & \deltapabar{} & Bar PA    & \amax{}    & [dp]  & \lbar{}   & [dp]      & \pabox{}  & \rbox{}   & [dp]  & $\rbox/\amax$ & $\rbox/\lbar$ \\
           & (\degr)       & (\degr)   & (\arcsec)  & (kpc) & (\arcsec) & (kpc)     & (\degr)   & (\arcsec) & (kpc) &               &               \\
(1)        & (2)           & (3)       & (4)        & (5)   & (6)       & (7)       & (8)       & (9)       & (10)  & (11)          & (12)          \\
\hline
NGC210     &  19 & 173  & 32.0  & 3.46 & 48.0 & 5.19 & \ldots      & \ldots      &  \ldots     &  \ldots     &  \ldots     \\
NGC450     &  51 &  42  & 13.0  & 1.37 & 17.0 & 1.80 & \ldots      & \ldots      &  \ldots     &  \ldots     &  \ldots     \\
NGC578     &  31 &  85  & 17.0  & 1.86 & 21.0 & 2.30 & \ldots      & \ldots      &  \ldots     &  \ldots     &  \ldots     \\
NGC615     &  11 & 162  & 22.0  & 2.68 & 26.0 & 3.17 & 159 & 11.0 & 1.34 & 0.50 & 0.42 \\
NGC918     &  46 &   6  & 6.5  & 0.64 & 8.3 & 0.82 & \ldots      & \ldots      &  \ldots     &  \ldots     &  \ldots     \\

\hline
\end{tabular}

\medskip

Bar and B/P bulge measurements for galaxies in the GoodPA subsample (barred galaxies with
$\deltapabar \le 60\degr$).
(1) Galaxy name. (2) \deltapabar{} = deprojected position angle between bar and disk major axis.
(3) Position angle of bar. (4) Observed semi-major axis of bar at maximum ellipticity.
(5) Deprojected size in kpc. (6) Observed upper limit on bar size. 
(7) Deprojected size in kpc. (8) Position angle of B/P bulge. (9) Observed radial size 
of B/P bulge. (10) Deprojected size in kpc. (11) (Deprojected) size of B/P bulge 
relative to bar, using \amax{} for bar size. (12) Same, but using \lbar{} for bar size.
The full table is available in the online version of this paper; we show a representative
sample here.

\end{minipage}
\end{table*}

\section{How Common Are B/P Bulges, and Which Galaxies Have Them?} 
\label{sec:trends}

In our GoodPA sample (84 barred galaxies), we find a total of 44 B/P
bulges, for an overall B/P fraction of $0.524\pm0.054$. We also find two galaxies
with bars currently in the buckling stage \citep{erwin-debattista16},
for a buckling fraction of $0.024^{+0.024}_{-0.012}$. This leaves us
with a ``thin-bar'' fraction of $0.452^{+0.055}_{-0.053}$.

However, our sample is clearly biased (by our diameter limit, if nothing
else) towards rather large, massive galaxies, so it is by no means clear
that this B/P fraction of $\sim 50$\% applies to all barred galaxies.
In the rest of this section, we investigate to what extent the presence
or absence of B/P bulges in bars depends on different galaxy parameters.
We are particularly interested in the gas mass fraction, since
at least some simulations have suggested that a high gas fraction in the
disc should suppress buckling (see references in Section~\ref{sec:data}).

\subsection{Trends with Galaxy Parameters: Logistic Regression}\label{sec:logistic} 

How the B/P fraction depends on different
parameters can be seen in
Figures~\ref{fig:fBP-mstar}--\ref{fig:fBP-amax}, where we plot B/P
fraction against galaxy stellar mass, gas mass fraction, Hubble type,
\bmv{} and \bmk{} colours, and bar size. We can see what appear to be
strong trends with \textit{all} the plotted parameters.
Figure~\ref{fig:fBP-mstar} shows that more massive galaxies are more
likely to have B/P bulges, with a strong transition at around $\logmstar
\sim 10.4$. Figure~\ref{fig:fBP-massratio} shows a somewhat weaker trend
with gas mass fraction, in that galaxies with high gas
fractions are \textit{less} likely to have B/P bulges -- or at least
B/P bulges are clearly less likely to be seen in galaxies with
$\fgas > 0.1$ and are not seen at all in galaxies with $\fgas \gtrsim 1$.
Figure~\ref{fig:fBP-Hubble-type} suggests that the B/P fraction is roughly
constant at around $\sim 80$\% for Hubble types S0--Sbc, and then drops
abruptly to very low values for types later than Sc.
Figure~\ref{fig:fBP-colors} shows strong trends with galaxy colour, in
the sense that redder galaxies are more likely to host B/P bulges.
Finally, Figure~\ref{fig:fBP-amax} shows that galaxies with small bars
are less likely to have B/P bulges, although the case for very large
bars is ambiguous.

To quantify the apparent trends -- and measure their statistical
significance -- we turn to logistic regression, which involves modeling
data with a function for the probability for a binomial property (in our
case, presence or absence of a B/P bulge):
\begin{equation} 
P(x) \, = \, \frac{1}{1 + e^{-\alpha + \beta x}}, 
\end{equation} 
where $P$ is the probability for a galaxy to have the particular
property and $x$ is the independent variable. The function ranges
between 0 and 1, as appropriate for modeling a probability. We use the
standard maximum-likelihood-based implementation in
R\footnote{\url{www.r-project.org}} to determine the best-fitting
parameters. (Note that logistic regression involves fitting \textit{all}
the individual data points; it is \textit{not} a fit to binned values,
such as those plotted in
Figures~\ref{fig:fBP-mstar}--\ref{fig:fBP-amax}, and thus is not biased
by any details of a particular binning scheme.) 

The R program also calculates the Akaike Information Criterion
\citep[AIC;][]{akaike74}:
\begin{equation}
\mathrm{AIC} \; = \; -2 \ln L \, - \, 2 k,
\end{equation}
where $L$ is the likelihood of the model and $k$ is the number of data points.
The AIC value has no meaning
in and of itself, but values from fits of different models to the same
data can be used in a model-comparison fashion, with lower values of AIC
indicating better relative agreement between model and data. In practice,
differences of $|\Delta{\rm AIC}| \lesssim
2$ are ignored, differences of 2--6 are weak evidence in favour of the
model with lower AIC, and differences $> 6$ are considered strong
evidence in favour of the model with lower AIC.

Table~\ref{tab:logistic} shows the result of our logistic fits for
\fBPbars{} -- i.e., the fraction of barred galaxies with B/P bulges -- as a
function of \logmstar, gas mass fraction \fgas, Hubble type $T$, \bmk{}
and \bmv{} colours, and both relative and absolute bar size (bar size as
a fraction of $R_{25}$ and bar size in kpc). The $P_{\beta}$ value is
the probability of obtaining a trend as strong as the observed one under
the null hypothesis of $\beta = 0$, and can be used as guide in
determining whether the trend is statistically significant or not. We
break the comparisons into different subsamples, depending on whether
the appropriate data is available for individual galaxies: all galaxies
(\logmstar, $T$, bar sizes), and subsamples for galaxies with detected
\hi{} and measured colours. (Because the \logmstar{} fit is by far the
best for the full sample, we also show \logmstar{} fits for the
limited-data subsamples, since AIC values can only be compared for fits
to the same data sets.)

What Table~\ref{tab:logistic} demonstrates is that the probability of
hosting a B/P bulge can be modeled with some success as a function of
\textit{each} of the galaxy parameters we tested. Given the visual
trends in Figures~\ref{fig:fBP-mstar}--\ref{fig:fBP-amax}, this
is not surprising, though the small values of $P_{\beta}$
($\lesssim 0.006$ in all cases) are evidence that these trends are probably not
statistical flukes. The AIC values suggest that \textit{best} individual
model, by a clear margin, is the dependence on stellar mass; this model is
plotted in  Figure~\ref{fig:fBP-mstar}. (We
also show what the fit looks like if we temporarily count the two
buckling-bar galaxies -- NGC~3227 and NGC~4569 -- as actually having B/P
bulges.) We can use this to define a
``transition mass'': the stellar mass for which the B/P fraction is
50\%. This happens for $\logmstar \approx 10.37$.

\begin{table}
\caption{Logistic Regression Results: Single Variables}
\label{tab:logistic}
\begin{tabular}{@{}lrrrr}
\hline
Variable & $\alpha$ & $\beta$  & $P_{\beta}$ & AIC \\
(1)      & (2)      & (3)      & (4)         & (5) \\
\hline
$\log \Mstar$   &  $-42.79$   & $4.13$   & $4.4 \times 10^{-6}$   & $72.26$ \\
Hubble type $T$   &  $1.94$   & $-0.51$   & $5.6 \times 10^{-5}$   & $96.58$ \\
\amax{} [kpc]   &  $-1.29$   & $0.48$   & $0.00039$   & $102.89$ \\
$\amax/R_{25}$   &  $-1.36$   & $6.44$   & $0.00086$   & $105.83$ \\
\lbar{} [kpc]   &  $-1.16$   & $0.33$   & $0.0017$   & $107.47$ \\
$\lbar/R_{25}$   &  $-1.10$   & $3.95$   & $0.0055$   & $111.05$ \\

\hline
\multicolumn{5}{c}{Subsample: galaxies with H~\textsc{i} detections} \\

$\log \Mstar$   &  $-51.54$   & $4.97$   & $8.6 \times 10^{-6}$   & $57.93$ \\
\fgas   &  $1.04$   & $-5.29$   & $0.0019$   & $86.99$ \\

\hline
\multicolumn{5}{c}{Subsample: galaxies with \bmv{} values} \\

$\log \Mstar$   &  $-51.69$   & $5.01$   & $1.6 \times 10^{-5}$   & $60.52$ \\
\bmv   &  $-7.64$   & $2.67$   & $2.1 \times 10^{-5}$   & $75.50$ \\

\hline
\multicolumn{5}{c}{Subsample: galaxies with \bmk{} values} \\

$\log \Mstar$   &  $-53.82$   & $5.21$   & $5.3 \times 10^{-6}$   & $61.04$ \\
\bmk   &  $-7.95$   & $2.74$   & $6.6 \times 10^{-6}$   & $78.68$ \\

\hline
\end{tabular}

\medskip

Results of single-variable logistic regressions: probability of a barred
galaxy having a B/P bulge as function of values of different parameters. (1)
Galaxy parameter used in fit (\Mstar{} = stellar mass; \fgas{} = gas
mass ratio; \amax{} = bar maximum-ellipticity radius in kpc;
$\amax/R_{25} =$ bar maximum-ellipticity radius relative to $R_{25}$;
\lbar{} = bar upper-limit radius in kpc; $\lbar/R_{25} =$ bar
upper-limit radius relative to $R_{25}$). (2) Intercept value for 
fit. (3) Slope for fit. (4) $P$-value for slope. (5) Akaike
Information Criterion value for fit; lower values indicate better
fits. The upper part of the table uses the full 84-galaxy GoodPA
subsample; the lower sections deal with specific subsamples.

\end{table}

\begin{table}
\caption{Logistic Regression Results: Multiple Variables}
\label{tab:multilogistic}
\begin{tabular}{@{}lrrrr}
\hline
Variable & $\alpha$ & $\beta$  & $P_{\beta}$ &  AIC \\
(1)      & (2)      & (3)      & (4)         & (5) \\
\hline
$\log \Mstar$   &  $-37.67$   & $3.68$   & $0.00013$   & $73.11$ \\
Hubble type $T$   &              & $-0.15$   & $0.3$   &           \\[2mm]
$\log \Mstar$   &  $-48.10$   & $4.74$   & $1.8 \times 10^{-5}$   & $73.04$ \\
$D_{25}$   &              & $-0.04$   & $0.27$   &           \\[2mm]
$\log \Mstar$   &  $-40.13$   & $3.83$   & $3.9 \times 10^{-5}$   & $73.35$ \\
$\amax/R_{25}$   &              & $1.96$   & $0.35$   &           \\[2mm]
$\log \Mstar$   &  $-46.11$   & $4.48$   & $8.9 \times 10^{-5}$   & $73.97$ \\
\amax{} [kpc]   &              & $-0.09$   & $0.59$   &           \\[2mm]
$\log \Mstar$   &  $-40.13$   & $3.83$   & $3.9 \times 10^{-5}$   & $73.35$ \\
$\lbar/R_{25}$   &              & $1.96$   & $0.35$   &           \\[2mm]
$\log \Mstar$   &  $-46.11$   & $4.48$   & $8.9 \times 10^{-5}$   & $73.97$ \\
\lbar{} [kpc]   &              & $-0.09$   & $0.59$   &           \\

\hline
\multicolumn{5}{c}{Subsample: galaxies with H~\textsc{i} detections} \\

$\log \Mstar$   &  $-39.54$   & $3.81$   & $0.00036$   & $70.04$ \\
\fgas   &              & $-0.41$   & $0.82$   &           \\

\hline
\multicolumn{5}{c}{Subsample: galaxies with \bmv{} values} \\

$\log \Mstar$   &  $-31.95$   & $2.67$   & $0.011$   & $69.27$ \\
\bmv   &              & $1.46$   & $0.053$   &           \\

\hline
\multicolumn{5}{c}{Subsample: galaxies with \bmk{} values} \\

$\log \Mstar$   &  $-34.44$   & $2.91$   & $0.0048$   & $70.19$ \\
\bmk   &              & $1.42$   & $0.056$   &           \\

\hline
\end{tabular}

\medskip

As for Table~\ref{tab:logistic}, but now showing results of multiple-variable logistic 
regressions: probability of a
barred galaxy having a B/P bulge as function of both \Mstar{} and a
second variable. (1) Galaxy parameter used in fit (see
Table~\ref{tab:logistic} caption). (2) Intercept value for fit. (3)
Slope for fit. (4) $P$-value for slope. (5) Akaike Information
Criterion value for fit.

\end{table}

\subsection{Multiple Variables and the Primacy of Stellar Mass}\label{sec:mstar-primacy} 

As noted above, we find evidence for trends in B/P-bulge frequency  with
\textit{all} the galaxy parameters we examined; the strongest such case
is for stellar mass. Since the other independent variables in
Table~\ref{tab:logistic} are generally known to be strongly correlated
with \Mstar{} \citep[e.g.,][for \fgas{} and \Mstar; see also
Figure~\ref{fig:params-vs-mstar}]{catinella10,catinella13}, it is worth
exploring whether the presence of B/P bulges depends on any of these properties
\textit{independently of} the \Mstar{} correlation. For example, if we
hold stellar mass fixed, does varying Hubble type or \fgas{} affect the
B/P fraction?

We explore these questions in three ways:
\begin{enumerate}
\item Logistic regression with multiple parameters;
\item Matched-pair analysis;
\item Comparative plots.
\end{enumerate}

\subsubsection{Multiple Logistic Regression} 

The logistic equation for multiple independent variables is a generalization of
the standard logistic equation:
\begin{equation} 
P \, = \, \frac{1}{1 + e^{-\alpha + \sum_{i} \beta_{i} x_{i}}},
\end{equation} 
where the $x_{i}$ are the different variables (e.g., stellar mass, colour, 
Hubble type, etc.).

The results of our two-variable logistic regression analysis are
displayed in Table~\ref{tab:multilogistic}. These show that in every
single case, stellar mass is the most -- or indeed \textit{only} --
significant variable. The second variable has a marginally statistically
significant effect only in case of colour (\bmv{} or \bmk). One can
also see that in every case, the AIC values for the multivariable fits
are either indistinguishable from or higher than the AIC values for
the single-variable fits (using the same subsamples) which use just \Mstar{}
(Table~\ref{tab:logistic}). In particular, the fit using just \Mstar{}
is favoured over fits using both \Mstar{} and colour (either \bmv{} or
\bmk), with $\Delta {\rm AIC} \sim 9$ for the two-variable fits using
colour.

We conclude that while there is \textit{perhaps} tentative evidence for
a dependence on galaxy colour (i.e., at the same stellar mass, redder
galaxies are more likely to have B/P bulges), the main result is the
clear dominance of stellar mass: \textit{whether a barred galaxy has a
B/P bulge or not is determined primarily -- indeed, almost entirely --
by its stellar mass.}

\subsubsection{Matched-Pair Analysis} 

\begin{figure}
\begin{center}
\hspace*{-1.5mm}\includegraphics[scale=0.43]{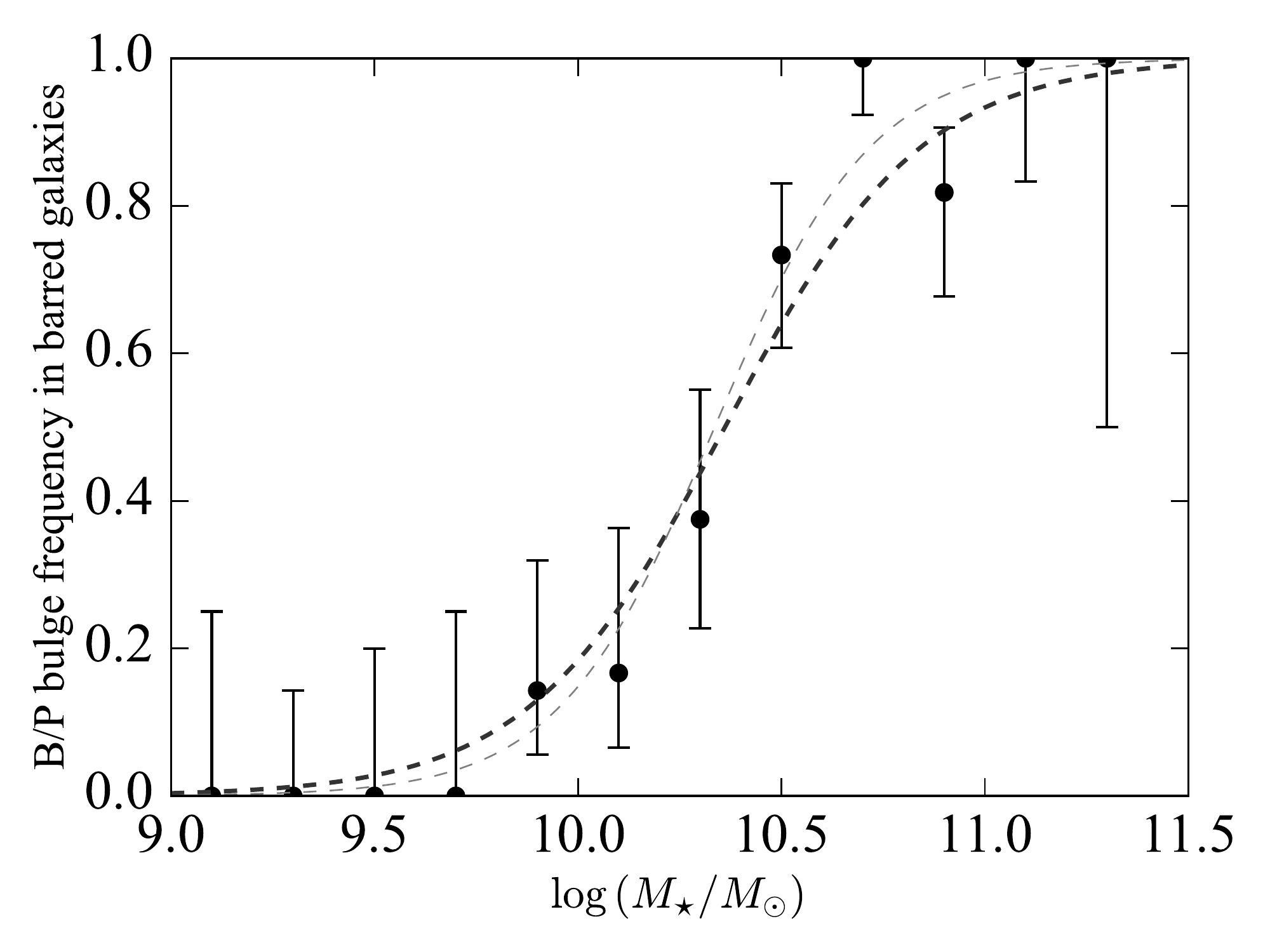}
\end{center}

\caption{Frequency of B/P bulges within bars as a function of galaxy
stellar mass. The thick dashed curve shows the best-fit logistic
regression (fit to the full set of individual data points rather than
the bins). The thin dashed curve shows the fit when considering the two
buckling-bar galaxies as having B/P bulges. \label{fig:fBP-mstar}}

\end{figure}

\begin{figure}
\begin{center}
\hspace*{-1.5mm}\includegraphics[scale=0.43]{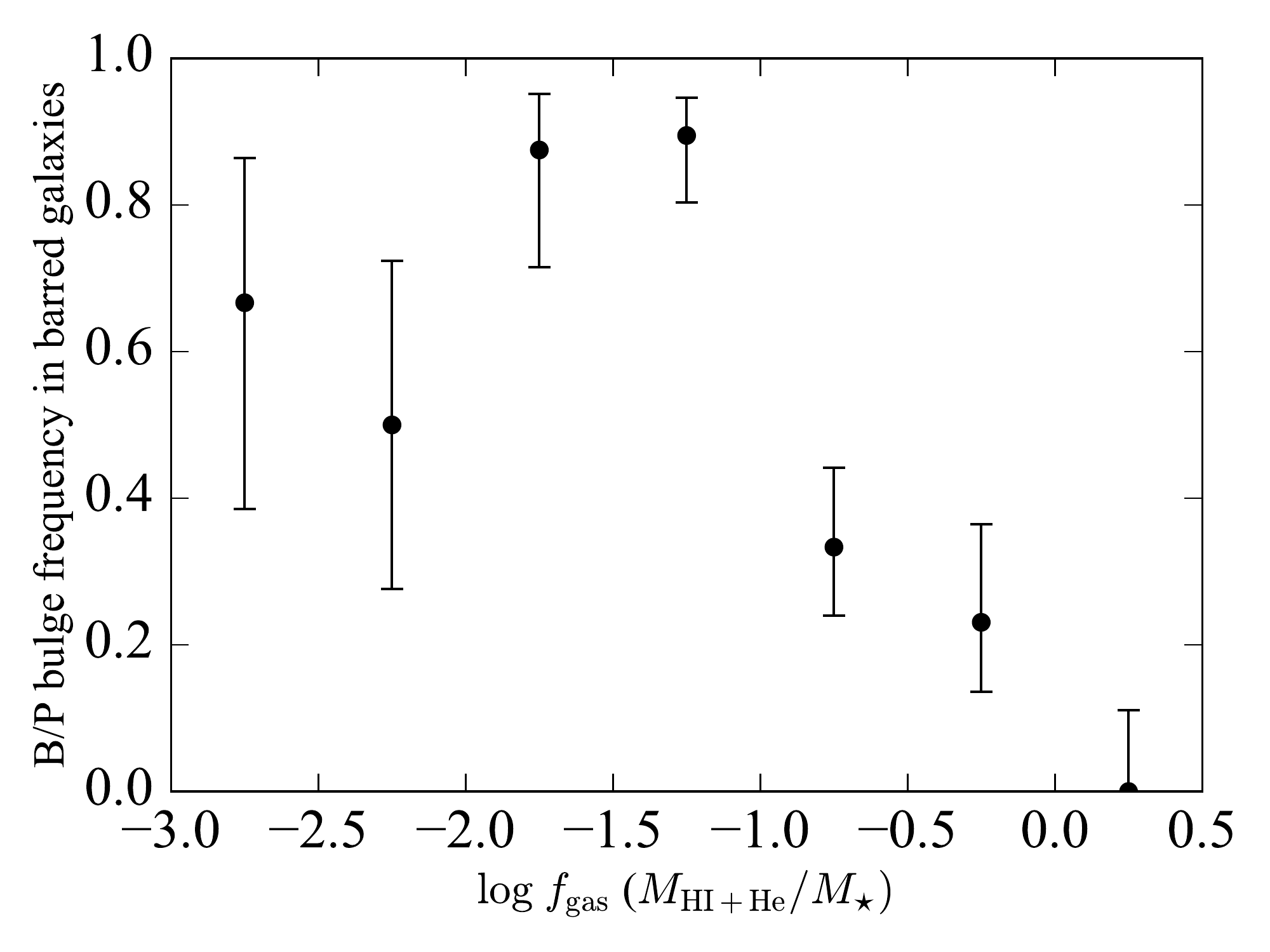}
\end{center}

\caption{Frequency of B/P bulges within bars as a function of gas mass
ratio \fgas{} ($= \Matomic/\Mstar$). \label{fig:fBP-massratio}}

\end{figure}

\begin{figure}
\begin{center}
\hspace*{-1.5mm}\includegraphics[scale=0.43]{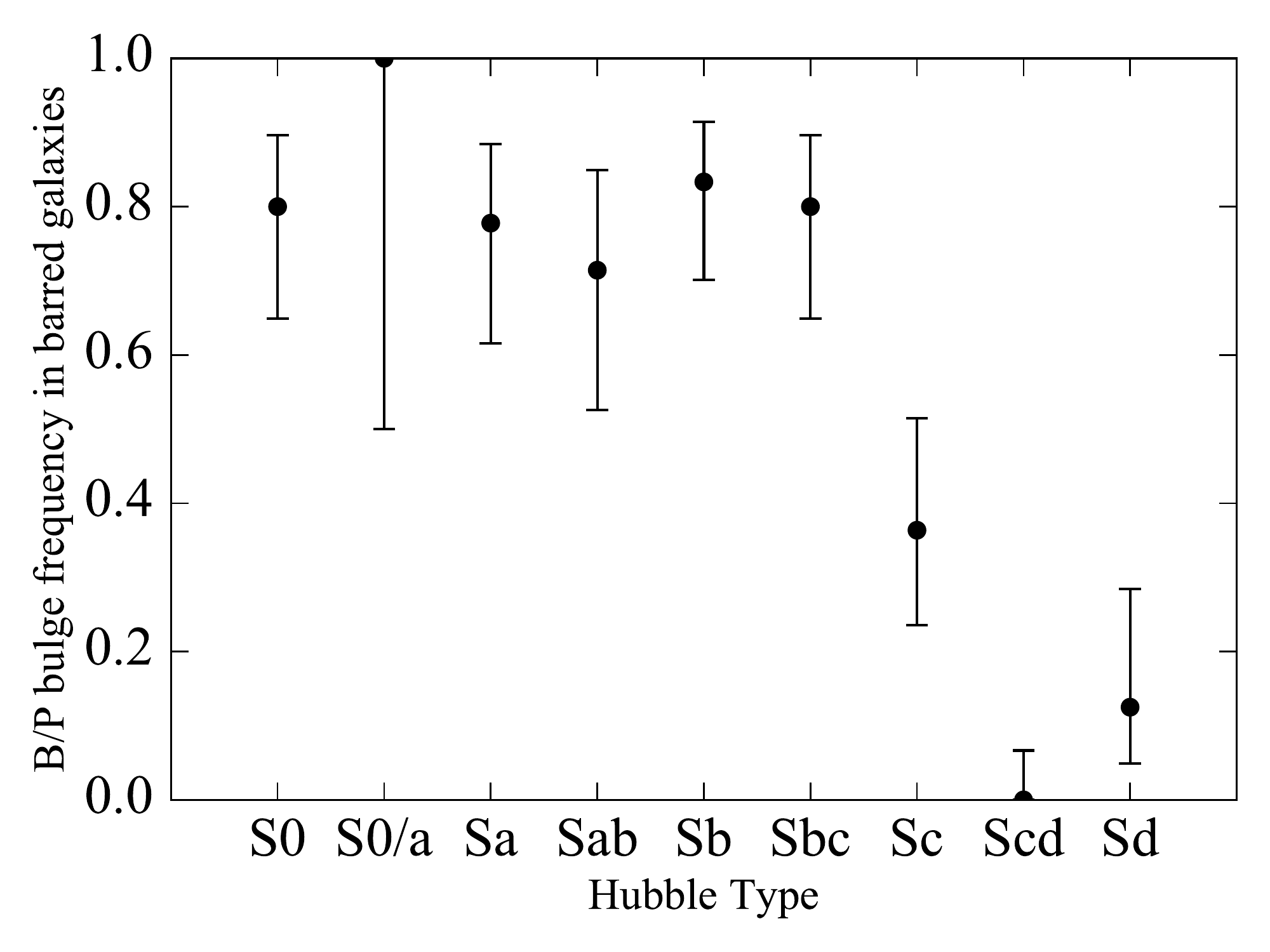}
\end{center}

\caption{Frequency of B/P bulges within bars as a function of Hubble type. 
\label{fig:fBP-Hubble-type}}

\end{figure}

\begin{figure*}
\begin{center}
\hspace*{-1.5mm}\includegraphics[scale=0.85]{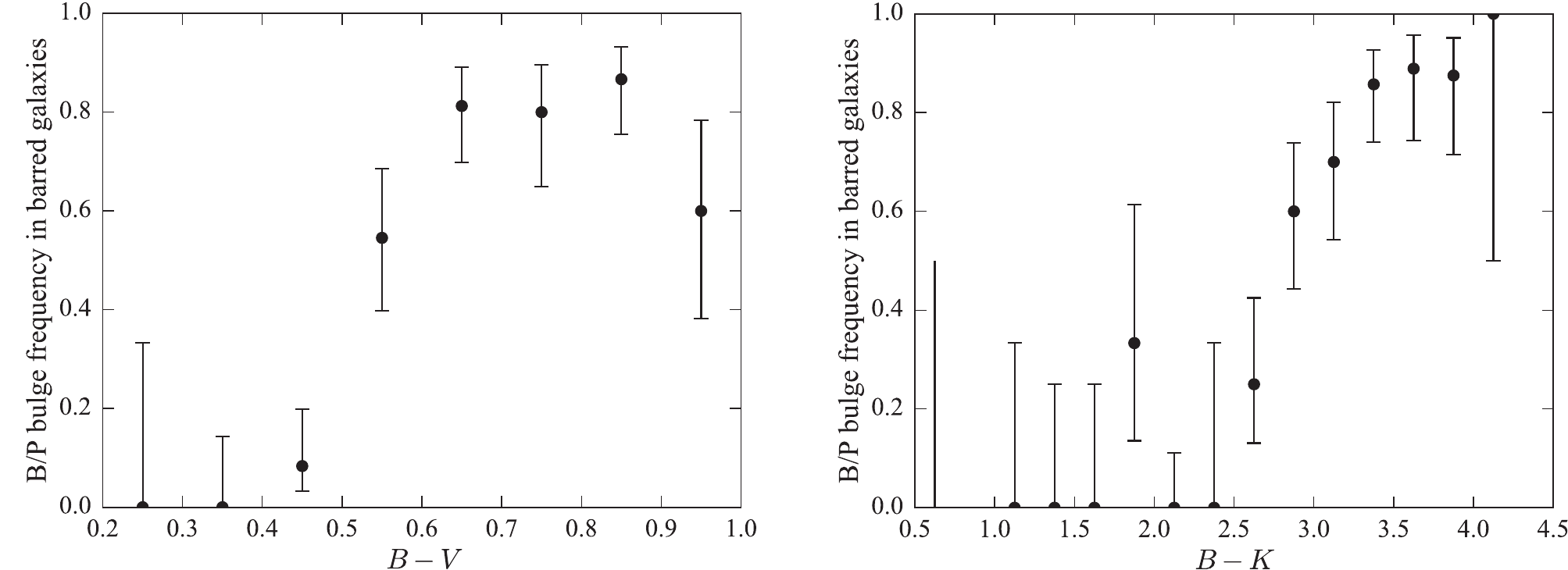}
\end{center}

\caption{Frequency of B/P bulges within bars as a function of galaxy colours:
\bmv{} (left) and \bmk{} (right).
\label{fig:fBP-colors}}

\end{figure*}

\begin{figure*}
\begin{center}
\hspace*{-1.5mm}\includegraphics[scale=0.85]{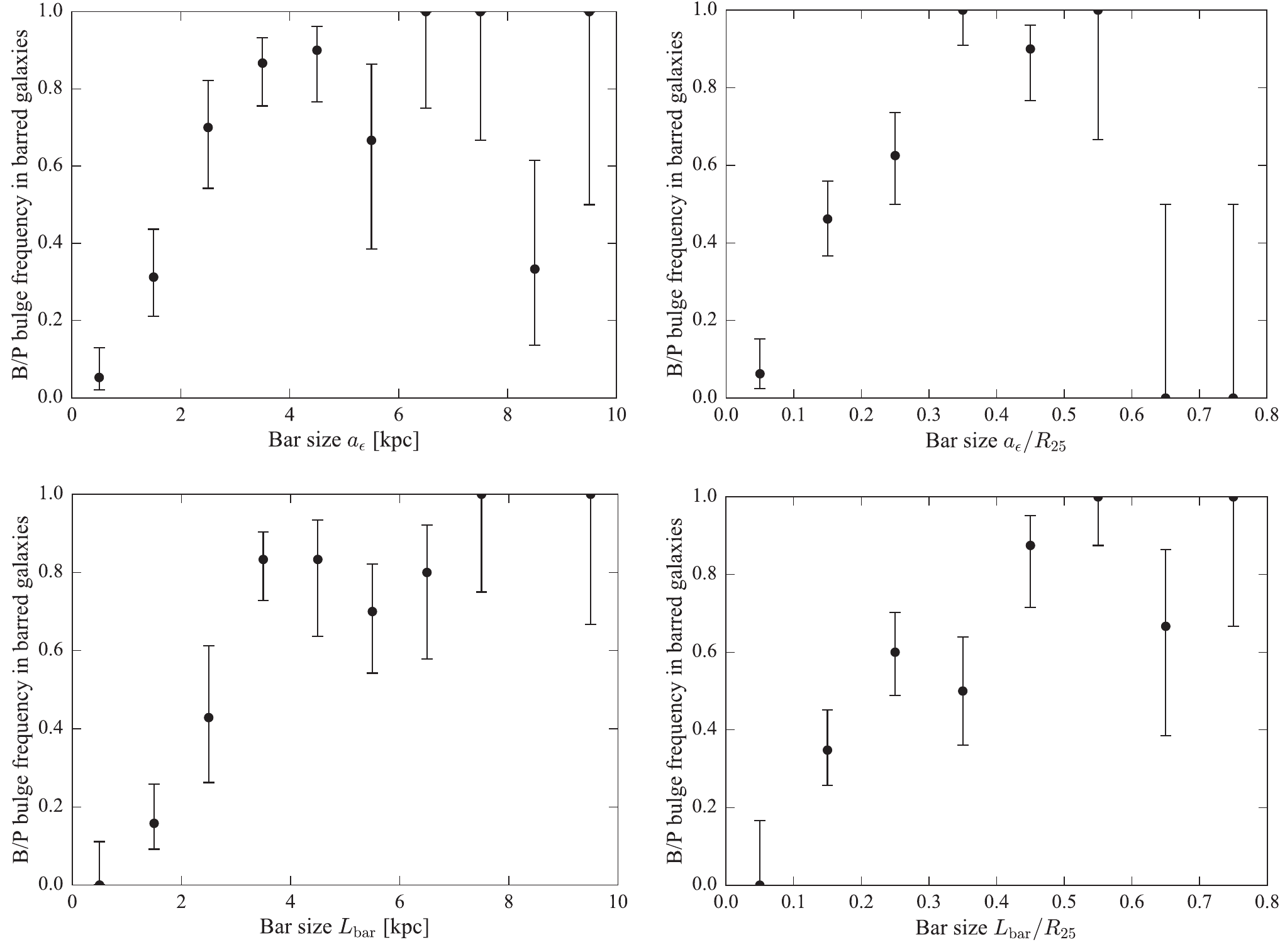}
\end{center}

\caption{Frequency of B/P bulges within bars versus absolute bar
size \amax{} in kpc (upper left), relative bar size $\amax/R_{25}$ (upper right),
absolute size \lbar{} (lower left), and relative size $\lbar/R_{25}$ (lower right).
\label{fig:fBP-amax}}

\end{figure*}

The limitation of logistic-regression analysis is that it assumes a
sigmoid relation between the probability of the binary variable and the
independent variable -- that is, $P(x)$ is either monotonically
increasing or monotonically decreasing with $x$. A more complicated
relation will not be well fit by the logistic function. Although
Figure~\ref{fig:fBP-mstar}--\ref{fig:fBP-amax} indicate that most of the
relations between B/P-bulge presence and variables such as \Mstar,
\fgas, colour, and Hubble type \textit{are} approximately monotonic, it
is worth exploring alternate, less model-dependent approaches to
finding evidence for secondary correlations.

A more model-independent way of testing for secondary correlations is
with a matched-pair analysis. This involves dividing a sample into two
subsamples based on the characteristic of interest (e.g., whether or not
the galaxy has a B/P bulge). Then each galaxy in the first subsample
(e.g., B/P-present) is paired with a random galaxy from the other
subsample (e.g., B/P-absent) which matches the first galaxy on some
particular parameter (e.g., stellar mass) within some tolerance. This is
repeated until all the galaxies in the first subsample have matches, or
until no more possible matches are left in the second subsample. (If a
randomly selected galaxy from the second subsample does not match within
the specified tolerance, another galaxy is selected. If this fails after
2000 attempts, the galaxy from the first subsample is discarded.) The
result is an approximation of matched test and control samples, with
similar distributions in the matching parameter. We can then compare
values of a second, ``comparison'' parameter (e.g., \fgas) between the
two subsample. We do this using the Kolmogorov-Smirnov and
Anderson-Darling two-sample tests. Because selecting random matching
pairs can introduce statistical fluctuations and thus produce
potentially misleading results, we repeat each analysis 200 times and
record the median values from the statistical tests (and also the median
values for the inter-pair differences in the comparison parameter.)

\begin{table}
\caption{Matched Pair Analysis: B/P vs non-B/P, Matched by Stellar Mass}
\label{tab:pairs}
\begin{tabular}{@{}lrrr}
\hline
Comparison & $\Delta$ & $P_{\rm KS}$  & $P_{\rm AD}$ \\
(1)      & (2)      & (3)           & (4) \\
\hline
\fgas   &  $-0.0103$   & $0.13$   & $0.13$ \\
Hubble type $T$   &  $-0.9000$   & $0.0022$   & $0.0026$ \\
\bmv   &  $0.0500$   & $0.029$   & $0.023$ \\
\bmk   &  $0.1280$   & $0.083$   & $0.046$ \\
$D_{25}$ [kpc]   &  $0.5464$   & $0.75$   & $0.7$ \\
$\amax/R_{25}$   &  $0.0457$   & $0.048$   & $0.019$ \\
\amax{} [kpc]   &  $0.8818$   & $0.007$   & $0.0045$ \\
$\lbar/R_{25}$   &  $0.0536$   & $0.048$   & $0.035$ \\
\lbar{} [kpc]   &  $0.8649$   & $0.0016$   & $0.0045$ \\

\hline
\end{tabular}

\medskip

Results of matched-pair analysis: B/P and non-B/P galaxies matched by
\Mstar. For columns 2--4, we show the median value from 200 rounds of matched-pair
analysis. (1) Comparison variable. (2) Difference in
comparison-variable value for paired galaxies (B/P $-$ non-B/P). (3) $P$-value from
Kolmogorov-Smirnov two-sample test. (4) $P$-value from Anderson-Darling
two-sample test.

\end{table}

Since our primary interest is in looking for possible trends apart from
the clear (and strongest) one with stellar mass, we ran the matched-pair
analysis by matching galaxies by stellar mass, with a tolerance of 0.1
dex in the log. The results of this analysis are summarized in
Table~\ref{tab:pairs}. Although many of the comparison parameters are
correlated with each other and thus not independent (e.g., \bmv{} and
\bmk{} colours are strongly correlated, as are bar sizes), we should
still be wary of the multiple-comparisons trap: the fact that running
multiple tests using different parameters increases the odds of
generating a ``statistically significant'' result purely by chance.
Accordingly, we suggest that the only genuinely significant results from
this analysis may be those for Hubble type and (absolute) bar size. 

Specifically, we find that for barred galaxies with similar masses, B/P
hosts tend to be about one Hubble stage earlier. Since Hubble type is
based partly on the degree of central concentration in a galaxy, and
since B/P bulges tend to be centrally concentrated structures
\citep[e.g.,][]{debattista04,athanassoula15}, it is possible that this
secondary Hubble-type correlation (earlier Hubble types are more likely
to host B/P bulges) is a side effect of B/P bulge formation: that is,
the presence of a B/P bulge biases the Hubble type classification to
earlier types. There may also be secondary correlations with absolute
bar length, in the sense that for galaxies of similar mass, the
B/P-bulge hosts will tend to have \textit{longer} bars (in the median,
the difference is slightly less than a kpc in both \amax{} and \lbar).

\subsubsection{Comparison Plots} 

\begin{figure*}
\begin{center}
\hspace*{-4mm}
\includegraphics[scale=0.92]{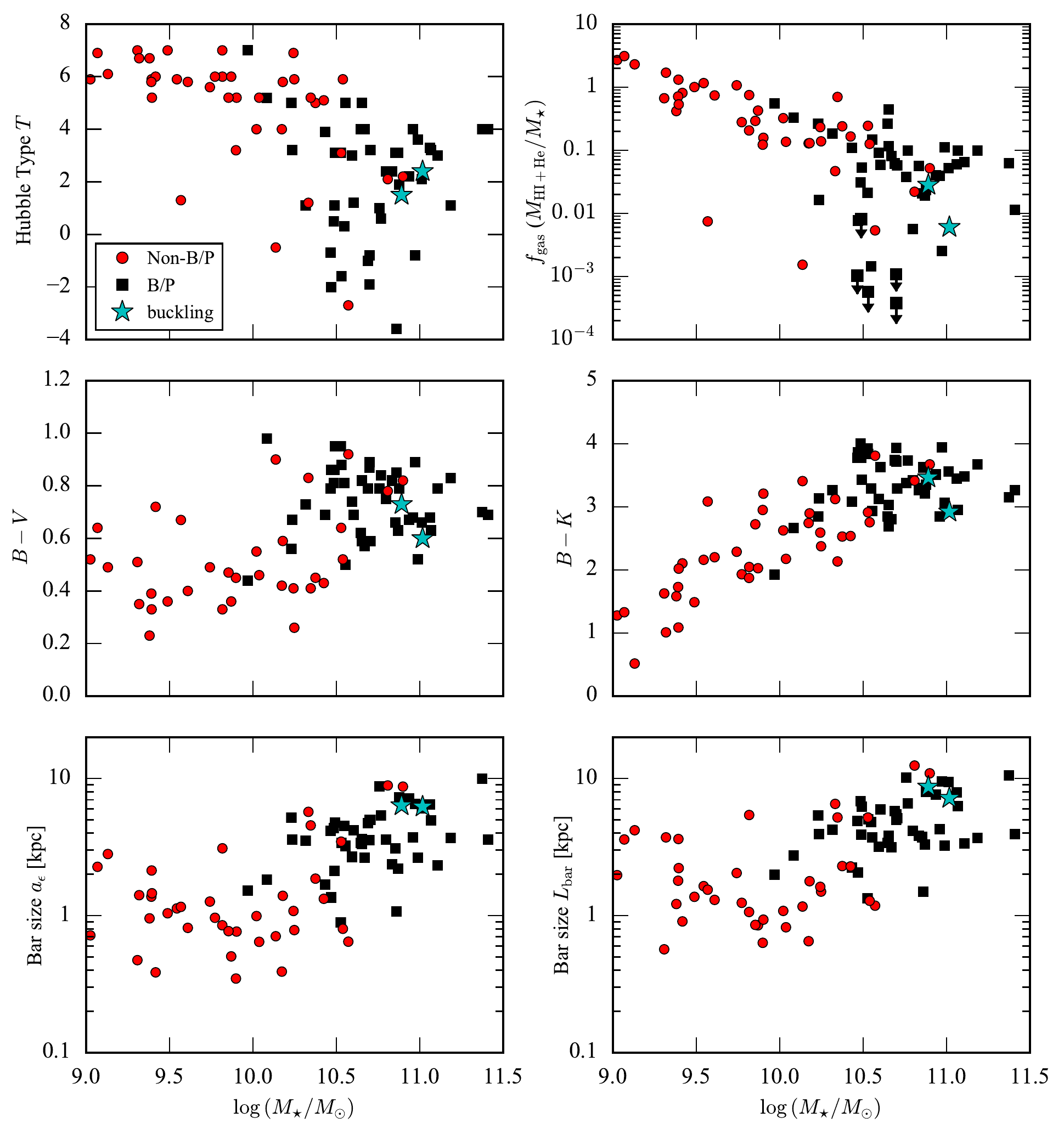}
\end{center}

\caption{Various barred-galaxy properties -- Hubble type $T$, gas mass
ratio (including five galaxies with \hi{} upper limits), colour, bar size
-- as a function of stellar mass for the GoodPA sample, showing galaxies
with B/P buges (solid black squares)] and without (red circles). Cyan
stars indicate the buckling bars in NGC~3227 and NGC~4569. In all cases,
the correlation between B/P state and stellar mass is the dominant one.
\label{fig:params-vs-mstar}}

\end{figure*}

In Figure~\ref{fig:params-vs-mstar} we plot different galaxy parameters
versus stellar mass for the GoodPA sample. We code the galaxies by
whether they have bars with B/P bulges (black squares) or whether they
have bars without them (red circles); we also show the two buckling bars
\citep{erwin-debattista16} using cyan stars. Although hints of weak
trends between the presence of B/P bulges and other parameters can
sometimes be seen -- e.g., for $\logmstar \sim 10.3$--10.5, galaxies
with B/P bulges tend to be redder than galaxies without -- it is clear
that the strong dependence of B/P presence on stellar mass dominates in
all plots.

\subsubsection{Summary} 

The clearest lesson from the various statistical (and graphical)
analyses in this section is that \textit{B/P bulge presence in barred galaxies
depends strongly -- indeed, almost entirely -- on the stellar mass of
the galaxy: more massive galaxies are more likely to have B/P bulges.} As
we have seen, there are also clear trends with, e.g., Hubble type, colour
(\bmv{} and \bmk), and bar size; but these can mostly if not entirely be
explained as side effects of the correlations of those parameters with
stellar mass. There is only weak evidence for additional,
\textit{independent} correlations of B/P presence with Hubble type
(earlier Hubble types are more likely to have B/P bulges) and possibly
with bar size (longer bars are more likely to have B/P bulges). Larger
samples with more galaxies in the stellar mass range of $\logmstar =
10$--10.5 are probably necessary in order to determine if these
secondary correlations are real.

\section{How Large Are B/P Bulges?}\label{sec:size} 

We measure the approximate sizes of B/P bulges as we did in
\citet{erwin-debattista13}: the half-length \rbox{} of the boxy region,
measured along its apparent major axis. (In practical terms, we measure
the full visible extent of the main boxy region on logarithmically
scaled images.)  By performing measurements on projections of $N$-body
simulations with orientations similar to those of real galaxies, we were
able to show in that paper that \rbox{} was a reasonable estimate for
the radial extent of the B/P bulge as measured from edge-on views of the
same simulations (see Fig.~8 of \citealt{erwin-debattista13}). The
measurements of \rbox{} and \pabox{} (the observed position angle of the
B/P major axis) are listed in Table~\ref{tab:bars}; we also list the
deprojected values of \rbox{} in kpc. Appendix~\ref{app:new-bp-plots}
shows isophote contours of individual galaxies with \rbox{} and \pabox{}
indicated.

By combining the sample in this paper with the measurements presented in
\citet{erwin-debattista13}, we find a total of nine galaxies which
\textit{also} have ``barlens'' sizes reported in \citet{laurikainen11};
as shown by \citet{laurikainen14} and \citet{athanassoula15}, the
barlens feature is another way of talking about the projected B/P bulge.
The mean size ratio of our measurements and theirs for the galaxies in
common is $\rbox/R(\mathrm{bl}) = 1.04 \pm 0.34$, where
$R(\mathrm{bl})$ is the radial size reported by \citet{laurikainen11}.
This is encouraging evidence that the two different approaches are
identifying and measuring the same structures, and that the different
measurements of B/P bulge size are not systematically biased (although
the large scatter indicates that measurements for individual galaxies
may not agree well).

On the other hand, a similar comparison for thirteen galaxies with
barlens measurements in \citet{herrera-endoqui15} yields
$\rbox/R(\mathrm{bl}) = 0.78 \pm 0.18$. It seems that the
Herrera-Endoqui et al.\ sizes are systematically larger than other
measurements: for 23 galaxies in common between
\citet{herrera-endoqui15} and \citet{laurikainen11}, the Laurikainen et
al.\ barlens sizes are generally smaller (mean size ratio $= 0.92 \pm
0.09$). 

\begin{figure}
\begin{center}
\hspace*{-15mm}
\includegraphics[scale=0.57]{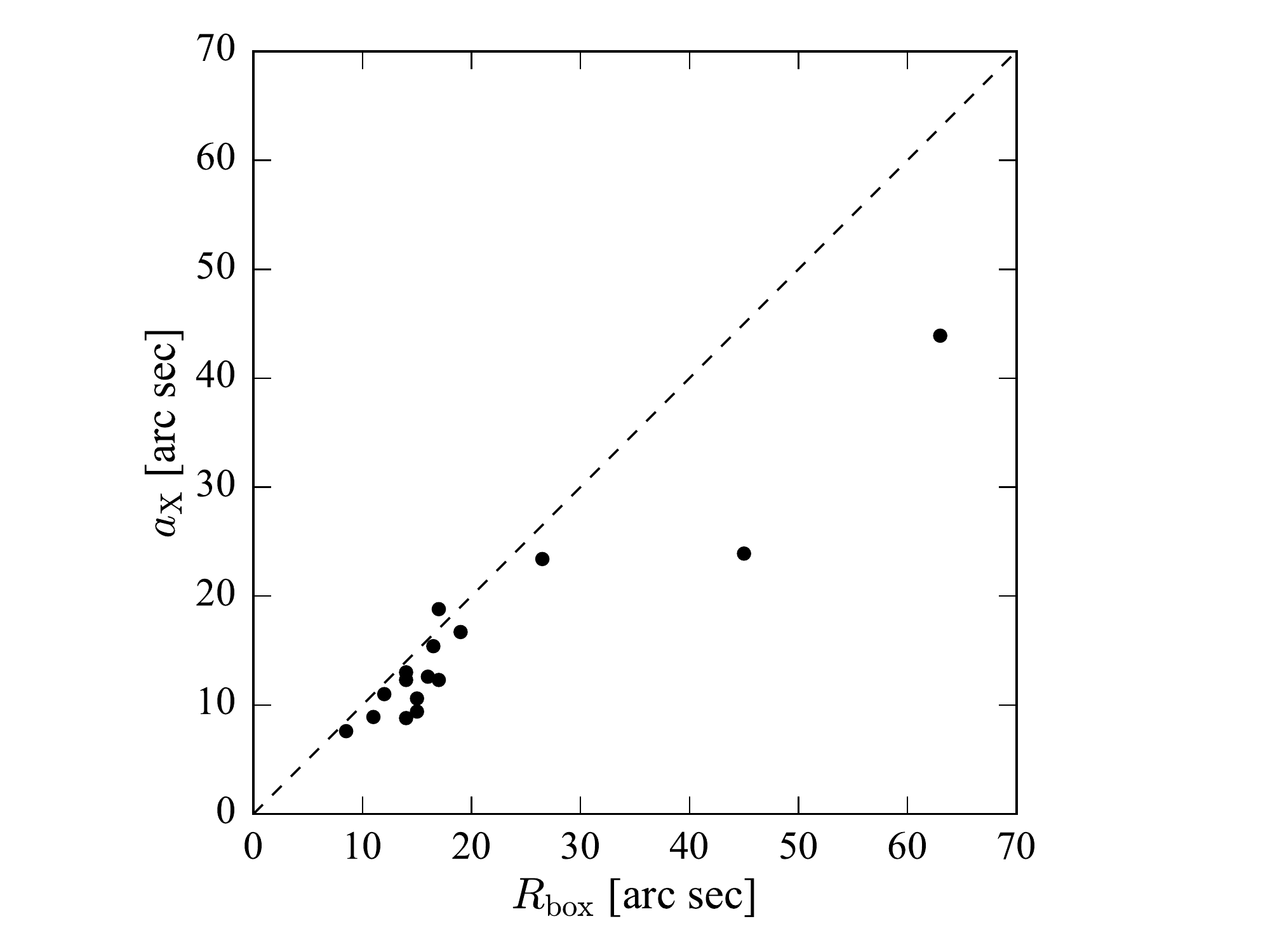}
\end{center}

\caption{Comparison of ``X-shape'' semi-major axis $a_{\mathrm{X}}$ from
\citet{laurikainen17} and our \rbox{} measurements for sixteen
galaxies. \label{fig:laurikanen16-comp}}

\end{figure}

Our $\rbox$ sizes are generally \textit{larger} than the ``X-shape''
sizes reported by \citet{laurikainen17} for sixteen galaxies which are
in our sample or in \citet{erwin-debattista13}, as shown in
Figure~\ref{fig:laurikanen16-comp}. The mean ratio is
$\rbox/a_{\mathrm{X}} = 1.28 \pm 0.25$, where $a_{\rm X}$ is the
measurement of the X-shape. We note that their X-shape sizes are also
systematically smaller than their barlens sizes (e.g., their Fig.~8).
Although they argue for a slight difference in Hubble types between
galaxies with barlenses and those with X-shapes, the difference between
our measurements and theirs for the same galaxies suggests that the
X-shape is on average systematically \textit{smaller} than the full B/P
bulge size, consistent with the idea that the X-shape represents
substructure within the overall B/P bulge rather than an alternative to
it.

In linear terms, \rbox{}  in our sample ranges from $\sim 450$~pc to
4.6~kpc, with a mean of $2.1 \pm 1.0$~kpc (median = 2.0 kpc). But a more
interesting question is probably: how large are B/P bulges relative to
the bars they live in (and formed out of)? This is important for
understanding the structure and underlying orbital skeletons of B/P
bulges, since different 3D orbit families can extend to different
fractions of the bar length
\citep[e.g.,][]{patsis02b,patsis06,portail15}.

\begin{figure}
\begin{center}
\hspace*{-4.5mm}\includegraphics[scale=0.45]{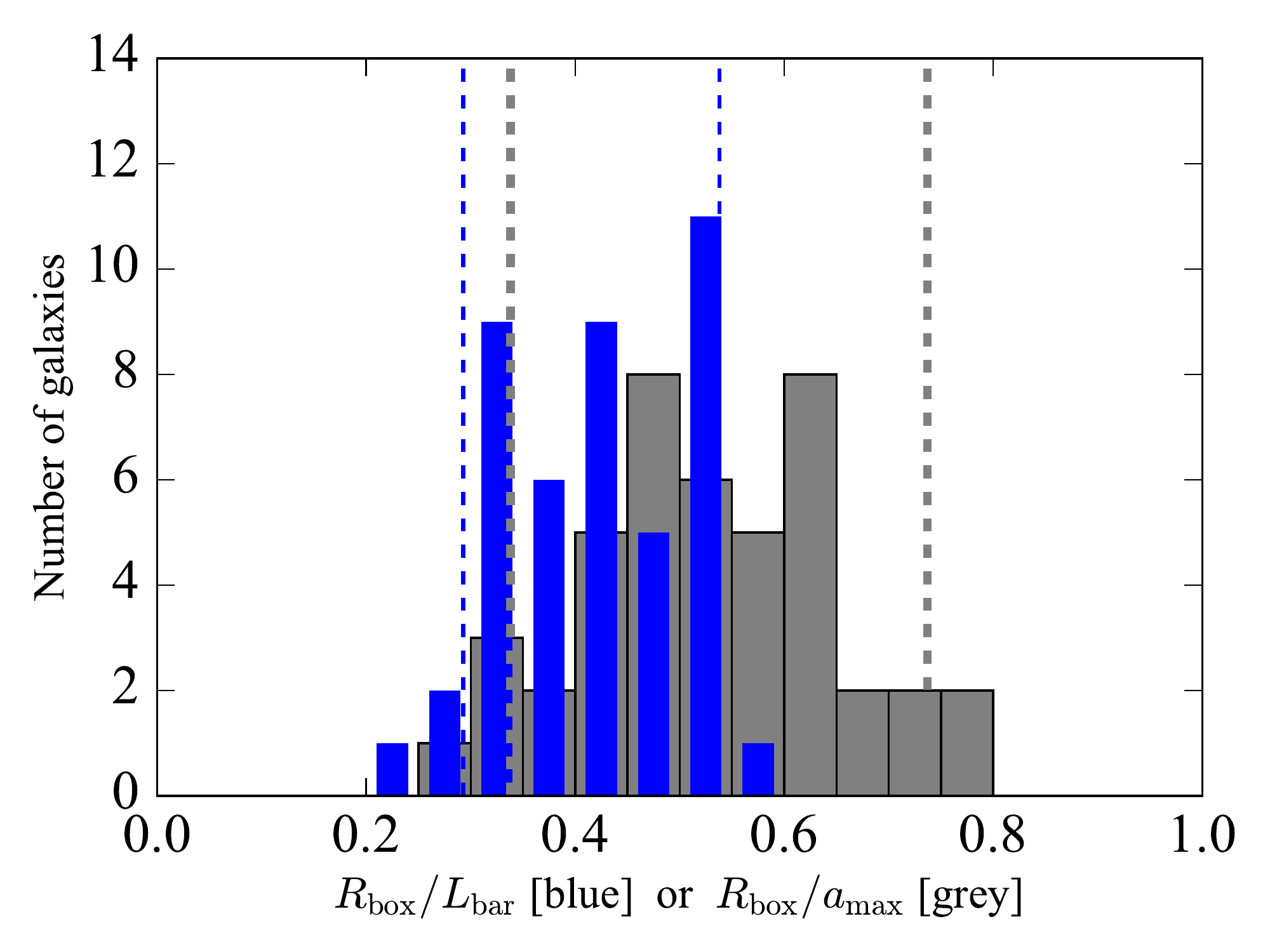}
\end{center}

\caption{Histogram of B/P bulge relative sizes: radius of B/P bulge \rbox{} relative to
either bar maximum-ellipticity semi-major axis \amax{} [grey] or full size
of bar \lbar{} [blue] for all barred galaxies in the GoodPA sample with
detected B/P bulges. Vertical dashed lines mark the relative sizes of buckling regions
in NGC~3227 (larger values) and NGC~4569 (smaller) values.\label{fig:rbox-Lbar-amax}}

\end{figure}

Columns 11 and 12 of Table~\ref{tab:bars} list the
(deprojected) relative sizes of the B/P bulges, using both of our bar
measurements (\amax{} and \lbar). Figure~\ref{fig:rbox-Lbar-amax} shows
histograms of B/P size relative to their host bars. The first histogram
(grey bars) uses the lower-limit \amax{} measurement of bar size; the
second (blue bars) uses the upper limit \lbar{} measurement, and is the
same as we used for a smaller sample of B/P bars in
\citet{erwin-debattista13}. The mean sizes are $\rbox/\amax = 0.53 \pm
0.12$ (median = 0.54) and $\rbox/\lbar = 0.42 \pm 0.09$ (median = 0.43),
respectively. This compares with mean $\rbox/\lbar = 0.43 \pm 0.10$
(median = 0.37) noted by \citet{erwin-debattista13} for their full
sample of 24 galaxies; their subset of 15 galaxies with $i > 40\degr$
and $\deltapabar < 45\degr$ had $\rbox/\lbar = 0.42 \pm 0.07$ (median =
0.43), which is essentially identical to our findings here. The larger
sample afforded by this paper\footnote{Note that the samples are not
entirely independent: six of the B/P-host galaxies in this paper's
sample are in the earlier sample of \citet{erwin-debattista13}.} does
allow us to see a broader spread in $\rbox/\lbar$, which can be as small
as 0.25 and as large as 0.76 (versus a maximum of 0.58 seen by Erwin \&
Debattista 2013).

As we pointed out in \citet{erwin-debattista13}, roughly
equivalent measurements based on near-IR imaging were reported for six
edge-on galaxies by \citet{lutticke00b}. The inversion of their
``BAL/BPL'' measurements yields $\rbox/\lbar = 0.38\pm0.06$, assuming
that their bar-length measurements are actually of bars and not, e.g.,
rings or lenses. This is entirely compatible with our relative sizes,
especially when the upper-limit bar size \lbar{} is used.

\citet{athanassoula15} reported mean relative B/P bulge sizes for 28
moderately inclined galaxies from \citet{laurikainen11}, where the B/P
bulges were identified from the ``barlens'' morphology. Their Fig.~6
shows a histogram of relative sizes ($R(\mathrm{bl})$ divided by bar
length), with an apparent median of $\sim 0.6$; this is consistent with
our $\rbox/\lbar$ measurements.\footnote{\citet{athanassoula15} do not
mention which of the three bar-size measurements reported for each
galaxy in \citet{laurikainen11} were used in determining the relative
sizes.}

Do the relative sizes of B/P bulges correlate with any galaxy
parameters? We have checked for possible correlations with the global
galaxy properties \Mstar, \fgas, Hubble type $T$, and colour, as well as
possible correlations with absolute and relative \textit{bar} size (bar
size in kpc or as a fraction of $R_{25}$). We find \textit{no} evidence
for clear correlations with \textit{any} of these properties, though
occasional individual correlations appear. For example, we find Spearman
$r_{s} = 0.34$ and $P = 0.022$ for $\rbox/\amax$ versus \bmk{} -- but since
this becomes $r = 0.06$ and $P = 0.69$ for $\rbox/\lbar$, we do not
consider this good evidence for a correlation between relative B/P size
and galaxy colour. We note that \citet{herrera-endoqui15} looked at the
relative sizes of barlenses in their \sfourg{}-based sample and found no 
significant correlation with stellar mass either.

The only hint of a possible correlation is with Hubble type, in the
sense that S0s appear to have slightly smaller relative B/P sizes than
spirals, with $P = 0.018$ from the Anderson-Darling two-sample test (but
only $P = 0.059$ for the K-S test). However, this correlation is only
significant when $\rbox/\amax$ is used; the difference mostly vanishes
when when $\rbox/\lbar$ is used instead (mean $\rbox/\lbar = 0.46$ for
S0 versus 0.42 for Sa--Sbc; Anderson-Darling $P = 0.28$). Since we have
tested nine different possible correlations for each of $\rbox/\amax$
and $\rbox/\lbar$, finding a marginally significant correlation purely
by chance is a serious possibility. Furthermore, Fig.~10 of
\citet{herrera-endoqui15} shows a slight \textit{increase} in relative
size for barlenses in S0s, which is additional evidence against there being a
significant decrease in relative B/P sizes for S0s. Our conclusion is
that there is no real evidence for a correlation between relative B/P
size and any of the galaxy properties we have considered.

\section{Discussion} 

\subsection{Comparisons with Edge-on Studies} 

Previous attempts to estimate the frequency of B/P bulges have focused
on edge-on galaxies, and have generally used photographic-plate data.
The earliest such attempt was probably that of \citet{jarvis86}, who
estimated an overall frequency of only 1.2\% for all disc galaxies.
However, as \citet{shaw87} pointed out, the majority of the sample was
small in angular size, and resolution effects probably imposed strong
limitations on the detectability of B/P bulges. 

\citet{shaw87}, using a sample of 117 visually edge-on galaxies with
diameters $D_{25} \ge 3.5\arcmin$, found that $\approx 20$\% of the
galaxies had evidence for B/P bulges; he noted that this was almost
certainly a lower limit, since some additional galaxies in the sample
had B/P structures which showed up in, e.g., unsharp masking but not in
the main analysis. He also found that B/P bulges were more common in
early and intermediate Hubble types (S0--Sbc). Using a larger (but not
diameter-limited) sample of 555 galaxies with axis ratios of $b/a < 0.5$
($i \ga 60\degr$), \citet{de-souza87} reported a B/P fraction of 13\%;
they found that the frequency was highest in S0 galaxies (33\%) and
lowest for Sc galaxies (3\%).

If we naively assume that our sample is similar (e.g., in stellar-mass
distribution) to the samples of \citet{shaw87} and \citet{de-souza87},
then we would predict a \textit{maximum} edge-on fraction of $\sim 33$\% (using
our bar fraction of 63\% and the total B/P fraction of 52\%). Given
that some fraction of edge-on B/P bulges will escape detection when the
bars are close to end-on, the results of \citet{shaw87} are plausibly
consistent with ours. Those of \citet{de-souza87} are not, but the fact
that they included many galaxies with intermediate inclinations and
small diameters may explain the lower frequency they found.  In both
cases, the evidence for higher fractions for earlier Hubble types is
consistent with our findings (e.g., Figure~\ref{fig:fBP-Hubble-type}). 

The largest photographic-plate analysis is that of \citet[][hereafter
L00]{lutticke00a}, who classified images of 1224 mostly edge-on S0--Sd
galaxies. For 734 galaxies they were able to classify the bulges into
four general types: ``peanut-shaped'' (type 1), ``box-shaped'' (type
2), ``close to box-shaped'' (type 3), and ``elliptical'' (type 4); most
of the unclassified bulges were cases of galaxies not close enough
to edge-on or else contaminated by bright stars. For the classified
bulges, they found that 45\% were types 1--3, which they considered
signatures of B/P bulges. Since some of their type 4 bulges would
presumably include B/P bulges in end-on bars, the real frequency would
be higher. (Similar results were found for a much smaller sample by
\citealt{dettmar90}.)

The B/P fraction reported by L00 is clearly higher than what we find.
The most dramatic difference is in the later Hubble types: L00 reported
B/P bulges at a roughly constant frequency as a function of Hubble type,
including 40\% for their Sd galaxies. This disagrees strongly with our
finding (and those of earlier studies) that B/P bulge frequency drops
sharply for late-type spirals: only $\sim 10$\% of the barred Sd
galaxies in our GoodPA subsample have B/P bulges, which implies a B/P
frequency of $\sim 8$\% for all Sd galaxies (given a bar fraction of
$\sim 81$\% for all Sd galaxies in our parent sample). This discrepancy
is difficult to explain, unless the L00 sample is somehow biased toward
high stellar masses, or else if there are systematic differences between
the edge-on and face-on Hubble-type classifications for later-type
spirals. Another possibility is that their high B/P fraction is
evidence for B/P-like structures in \textit{unbarred} galaxies, as
suggested by \citet{patsis02a}.

The most recent attempt to estimate B/P-bulge fractions in edge-on
galaxies is that of \citet{yoshino15}, who identified B/P bulges by
performing 2D S\'ersic + exponential image decompositions of $\sim 1300$
edge-on galaxies using SDSS DR7\footnote{Sloan Digital Sky Survey Data
Release 7 \citep{abazajian09}.} images and then examined the residuals
for patterns characteristic of  a boxy or peanut-shaped bulge. (They
carried out a somewhat similar analysis on a separate sample of face-on
galaxies in order to estimate the overall bar fraction.) They found a
B/P-bulge fraction of $22 \pm 1$\% for edge-on galaxies in the $i$-band.
This is almost identical to the \citet{shaw87} result, and is plausibly
consistent with our results, but disagrees with the high fraction of
L00. Yoshino \& Yamauchi also concluded that about half of bars have B/P
bulges, which is in fact what we find for our sample.

\subsection{Evidence for the B/P Bulge--\Mstar{} Correlation in
Data from Other Studies} 

We have found a strong correlation between the presence of B/P bulges in
barred galaxies and the stellar mass of the galaxy
(Section~\ref{sec:trends}). Is there any evidence for this correlation
in edge-on galaxies? We explore this question by looking at the largest
and most recent edge-on study, that of \citet{yoshino15}, which has the
advantage of having reasonably accurate stellar masses available for
almost all their galaxies.

We started with their $i$-band edge-on subsample and restricted it to
galaxies with redshifts between 0.01 and 0.1 (we chose the $i$-band
because, as Yoshino \& Yamauchi note, it minizes dust obscuration, which
is especially important for edge-on galaxies); this yielded a sample of
1244 edge-on galaxies. We then matched these with the photometry-based
stellar masses for the same galaxies from the MPA-JHU DR7
database\footnote{http://wwwmpa.mpa-garching.mpg.de/SDSS/DR7/}. These
masses are based on SED fits to DR7 photometry following the
approach of \citet{salim07}; we used the ``median'' stellar-mass values
for each galaxy.

\begin{figure*}
\begin{center}
\hspace*{-2.5mm}
\includegraphics[scale=0.87]{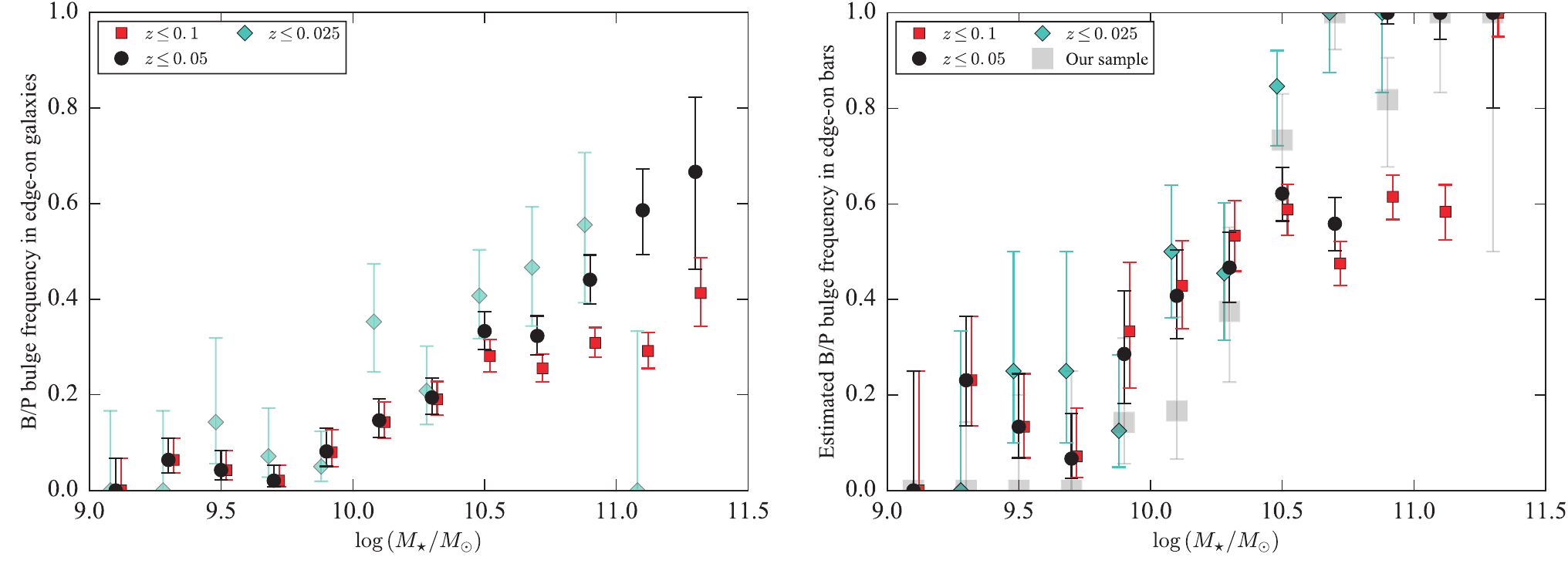}
\end{center}

\caption{Left: Fraction of edge-on galaxies in \citet{yoshino15} sample
(galaxies with $z = 0.01$--0.1) with detected B/P bulges as a function
of stellar mass. Three different upper redshift limits are shown (red
squares = $z \leq 0.1$, black circles = $z \leq 0.05$, cyan diamonds =
$z \leq 0.025$). Right: Estimated fraction of edge-on bars with B/P
bulges in \citeauthor {yoshino15} sample, using their face-on sample to
estimate bar fractions as a function of stellar mass. Also plotted for comparison
are the directly measured fractions from the analysis of our 84
moderately inclined GoodPA galaxies, using light grey squares (Section~\ref{sec:logistic};
Figure~\ref{fig:fBP-mstar}).\label{fig:yoshino}}

\end{figure*}

The left-hand panel of Figure~\ref{fig:yoshino} shows the frequency of
B/P bulges in edge-on galaxies in \citet{yoshino15} as a function of
stellar mass. Since more distant galaxies will, on average, have
structures smaller in angular size and thus more difficult to detect, we
plot the results using three different upper limits on redshift ($z \le
0.1$, 0.05, and 0.025) to see what effect changing the effective
resolution has (the resulting subsample sizes are 1244, 800, and 150,
respectively). In all three cases, we see a strong trend: B/P bulges are
more common in higher-mass galaxies, just as we have seen for our local
sample. Restricting the subsamples to lower redshifts increases the
detected frequencies, as we might expect, but does not change the basic
trend.

We cannot compare these B/P frequencies with ours directly because ours
are fractions for \textit{barred} galaxies \fBPbars, while those for the
\citet{yoshino15} data are for all (edge-on) galaxies. However, we
\textit{can} make a crude estimate of what \fBPbars{} would be for the
\citeauthor{yoshino15} data by using their $i$-band \textit{face-on}
sample to determine the bar fraction as a function of stellar mass
(using the same MPA-JHU source above), and then using these fractions to
estimate the total number of bars in each stellar-mass bin of the
edge-on sample. The result is shown in the right-hand panel of
Figure~\ref{fig:yoshino}; we include the direct measurements of
\fBPbars{} from \textit{our} data for comparison.\footnote{Since we do
not know which individual galaxies in the \citeauthor{yoshino15} edge-on
sample are barred-but-not-B/P, we cannot perform a logistic regression
analysis as we did for our sample.} The general trend is similar to that
for our data, and is rather close in some places (especially for the $z
\leq 0.05$ subsample); the transitional stellar mass, where $\fBPbars
\approx 50$\%, is $\logmstar \sim 10.3$--10.4 in all cases.

As we have noted previously, the B/P fraction measured for edge-on
galaxies is expected to be lower than what we find for our GoodPA
sample, due to the fact that some bars in edge-on systems will be
oriented close to end-on and thus have B/P bulges that appear round like
classical bulges, rather than boxy or peanut-shaped. But for several
mass bins in the right-hand panel of Figure~\ref{fig:yoshino}, the
estimated B/P fractions for the \citet{yoshino15} edge-on sample is
\textit{higher} than our fractions. 

\begin{figure}
\begin{center}
\hspace*{-3mm}
\includegraphics[scale=0.44]{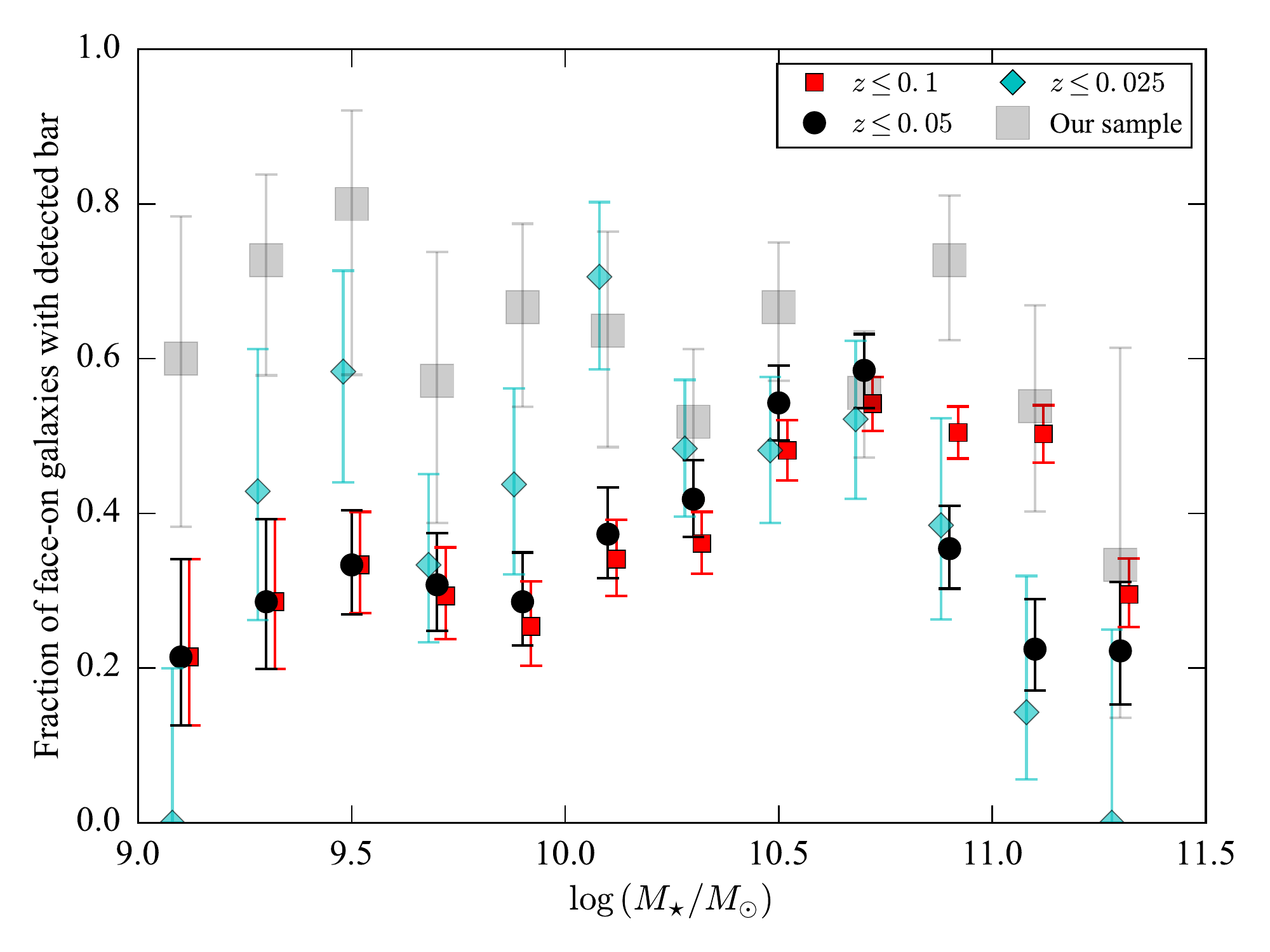}
\end{center}

\caption{Bar fractions for face-on galaxies in \citet{yoshino15} as a
function of stellar mass; symbols and colours as for
Figure~\ref{fig:yoshino}. Also shown are the bar fractions for the 186
galaxies of our Parent sample (light grey squares); the generally higher
fractions for our data are most likely a result of the use of near-IR
images and better spatial
resolution, since all our galaxies are at $z <
0.01$.\label{fig:yoshino2}}

\end{figure}

One possible reason for this discrepancy is that the \textit{face-on}
bar fraction in the \citet{yoshino15} data -- which is the denominator
for calculating \fBPbars{} from their data -- could be underestimated.
This possibility stems from the fact that the SDSS images have only
moderate resolution (FWHM typically $\sim 1.5\arcsec$) and, more
importantly, the fact that most of the galaxies are relatively far away,
which limits the effective spatial resolution. For example, the median
redshifts for the three subsamples of their data that we use are 0.021,
0.033, and 0.042, for the $z < 0.025$, $z < 0.05$, and $z < 0.01$
subsamples, respectively. For a typical SDSS seeing of $\sim
1.5\arcsec$, this translates to spatial resolution limits of $\sim
400$--850 pc; the practical limit for bar semi-major axes will probably
be several times larger. In contrast, for the median distance of our
sample (16.9 Mpc), the \textit{Spitzer} 3.6\micron{} FWHM of $\sim
1.6\arcsec$ translates to 130 pc. An additional factor is the fact that
our bar and B/P-bulge detections use near-IR images (mostly 3.6\micron),
which enhances the ability to detect bars compared to the optical SDSS
image \citep[e.g.,][]{eskridge00}.

Figure~\ref{fig:yoshino2} shows the detected bar fractions for the
\citet{yoshino15} face-on sample as a function of stellar mass (with the
same three redshift cutoffs), along with the bar fraction for our
complete Parent sample. As expected, our detected fractions are
consistently higher. This supports our suspicion that the values
of \fBPbars{} that we have estimated for the
\citet{yoshino15} sample are probably too high, because
the face-on \textit{barred} fractions are probably too low.

Finally, we note that the study of \sfourg{} galaxies by
\citet{herrera-endoqui15} -- based on the classifications of
\citet{buta15} -- found barlenses (analogous to the box/oval zone we
identify with B/P bulges) almost exclusively in galaxies with stellar
masses $> 10^{10} \Msun$ (see their Fig.~10), which is consistent with
our trend. (They also identify barlenses only in galaxies with Hubble
types of Sbc and earlier, which matches our finding as well;
Figure~\ref{fig:fBP-Hubble-type}).

The data of \citet{yoshino15} and, to a lesser extent, of
\citet{herrera-endoqui15} thus provide independent confirmations of our
basic finding: B/P bulge frequency is a strong function of stellar mass.

\subsection{The Dependence -- or Lack Thereof -- of B/P Fraction on Galaxy Properties: Gas Fraction
and Stellar Mass} 
\label{sec:discussion-properties}

Simulations have suggested that a significant gas mass fraction in the
disc can weaken, delay, or suppress the buckling of bars
\citep{berentzen98,debattista06,berentzen07,wozniak09}. The actual
relative mass fraction -- that is,  $\fgas = M_{\rm gas}/\Mstar$ -- can
be rather low. For example, \citet{berentzen07} found suppression of
buckling for (constant) values of $\fgas \ga 0.03$, and buckling was
suppressed in the simulation of \citet{wozniak09}, during which \fgas{}
declined from 0.11 to 0.047 due to the transformation of gas into
stars.  In our data we find no evidence for a correlation between
(atomic) gas mass fraction and B/P presence or absence -- once the
correlation between gas content and stellar mass is controlled for, that
is. Moreover, Figures~\ref{fig:fBP-massratio} and
\ref{fig:params-vs-mstar} show that B/P bulges remain fairly common for
values of \fgas{} as high as $\sim 0.2$--0.3 -- well above the level
explored in most of the simulations. (Recall that our \fgas{} does
\textit{not} include contributions from molecular gas, so our values are
really lower limits.)

At first glance, our results appear to contradict the theoretical
work. However, there are two reasons why we might not actually expect a strong
correlation between B/P presence and gas content. 

The first point is that we are using measurements of the
\textit{current} gas-mass fraction in our sample. If most (or all) of
the B/P bulges in our sample arose from buckling events, most of these
events would likely have occurred several Gyr ago, when the gas mass
ratios were probably different. Unfortunately, this is not something
that we can measure for our galaxies. (It is more
likely that these galaxies had, on average, \textit{higher} gas
fractions in the past, which would make the problem worse.) A more
comprehensive test of whether gas mass fraction affects the formation of
B/P bulges by buckling will probably have to come from detailed
observations at higher redshifts, when the buckling frequency would have
been at its peak \citep{erwin-debattista16}.

The second point is that the same gas-rich simulations which showed suppression
of \textit{buckling} generally also showed \textit{symmetric} B/P
growth, so that the bars often ended up with strong B/P bulges anyway.
Thus, despite our recent demonstration that buckling does indeed take
place in at least some real galaxies \citep{erwin-debattista16}, the
presence of B/P bulges in galaxies with $\fgas \ga 0.05$
could be an indication that most B/P bulges actually result from
\textit{symmetric}, non-buckling growth.

On the other hand, as our analysis in Section~\ref{sec:logistic} and
Figure~\ref{fig:params-vs-mstar} shows, gas mass fraction is much less
important than stellar mass in determining whether a barred galaxy has a
B/P bulge. So while it could be that the gas content helps determine
\textit{how} a B/P bulge forms -- via buckling or via symmetric growth
-- \textit{whether} a B/P bulge forms at all seem rather independent of
the (current) gas mass fraction.

The strong dependence of B/P fraction on galaxy stellar mass is somewhat
surprising, since we are aware of no theoretical predictions that would
warrant such a trend. As noted above (Section~\ref{sec:mstar-primacy}),
variables which might have been thought to affect B/P presence -- gas
mass fraction, or perhaps colour as a proxy for mean stellar age -- do
not show much evidence for secondary correlations with B/P fraction, and
their correlation with B/P fraction can be explained as a side-effect of
their own correlations with stellar mass. One could speculate for a
possible correlation between current stellar mass and gas mass fraction
\textit{at the time when the bar formed}, but is difficult to see how to
model this, let alone how to test this.

One possible connection might be with the (indirect) evidence for
relatively higher vertical velocity dispersions in lower-mass galaxies.
For example, comparisons of vertical and radial disc scale lengths
indicate that lower-mass galaxies tend to have relatively thicker discs
\citep[e.g., Fig.~1 in][]{bershady10}; this implies that the ratio of
vertical to radial stellar velocity dispersion $\sigma_{z}/\sigma_{R}$
is likely higher in lower-mass galaxies. This is relevant to the
question of when and whether bars buckle, because the buckling instability is
thought to happen when $\sigma_{z}/\sigma_{R}$ drops below some
critical threshold, due to the increase in $\sigma_{R}$ that follows bar
formation \citep[e.g.,][]{toomre66,raha91,merritt94,martinez-valpuesta06}. If
$\sigma_{z}$ is higher to begin with, then it could be harder for
buckling to happen in lower-mass galaxies. Unfortunately, direct
measurements of both $\sigma_{z}$ and $\sigma_{R}$ in disc galaxies are
rare, so it is difficult to test this idea.

\subsection{Current Buckling in Bars} 

In \citet{erwin-debattista16}, we presented evidence indicating that the
bars in NGC~3227 and NGC~4569 -- both of which are in our GoodPA
sample -- are \textit{currently} buckling. Given the large
size of both the bar and the buckling region in NGC~3227 ($\lbar =
8.7$~kpc deprojected, $\rbox/\lbar \approx 0.54$), it might be possible
that it is undergoing a \textit{secondary} buckling event
\citep{martinez-valpuesta06}, but this seems very implausible for
NGC~4569, where the buckling is confined to the inner part
of the bar ($\rbox/\lbar = 0.29$).

Both galaxies are quite massive ($\logmstar = 10.89$ and 11.02 for
NGC~3227 and NGC~4569, respectively); Figure~\ref{fig:fBP-mstar}
shows that this is in the regime where $\sim 80$\% of barred galaxies
already have B/P bulges. This suggests two interesting questions: why
have these galaxies not buckled previously (a question which of course
applies to the other massive barred galaxies without B/P bulges), and
why are they buckling \textit{now}? One possibility is that the bars
have only formed relatively recently in each galaxy.

\subsection{The B/P Bulge of the Milky Way in Context} 

As noted in the Introduction, our own Galaxy has a bar with a B/P bulge.
Is this unusual, or is it something to be expected given other
attributes of the Galaxy? Since most estimates for its stellar mass lie
in the range 5--$6.5 \times 10^{10} \Msun$
\citep[e.g.,][]{flynn06,mcmillan11,licquia15,mcmillan16}, the logistic
fit of Section~\ref{sec:logistic} for \fBPbars{} as a function of the
stellar mass would predict probabilities of 0.79--0.86
for the Galaxy's bar to host a B/P bulge; the observed frequency for
that mass range is 0.88.  This suggests that our galaxy is quite typical
in having a B/P bulge.

\citet{wegg15} projected their model of the central Galaxy as if it were
seen with $\deltapabar = 45\degr$ and an inclination of 60\degr{}, and
then used the approach of \citet{erwin-debattista13} to derive a
relative B/P size of $\rbox/\lbar \approx 0.26$. In
Section~\ref{sec:size}, we noted that $\rbox/\lbar = 0.42 \pm 0.09$ for
the galaxies in our sample, with a range of 0.25 to 0.76. The Galaxy's
B/P is thus very near the lower end of the range of observed relative
B/P sizes, though not outside it.

Given the mass of the Galaxy and its barred nature, we can say that it
would if anything be \textit{unusual} for the Galaxy to \textit{not}
have a B/P bulge. The relatively small size of the B/P bulge compared to
the full, in-plane length of the bar is, however, somewhat unusual.

\section{Summary} 

From a parent sample of 186 local disc galaxies with inclinations
between 40\degr{} and 70\degr, we have carefully defined a subset of 84
barred galaxies with orientations (bar position angle $\leq 60\degr$
away from the disc major axis) ideal for determining the presence or
absence of boxy/peanut-shaped (B/P) bulges inside the bars. Within this
subsample, we find a total of 44 barred galaxies with B/P bulges, plus
two more that are currently in the vertical buckling stage and will
likely transform into hosts of B/P bulges in the next Gyr or so. The
remainder show no evidence for B/P bulges and most likely have
vertically thin bars.

Extensive statistical tests, including multiple logistic regression and
matched-pair analysis, show that the dominant correlation for B/P bulge
frequency is with \textit{stellar mass}, in the sense that more massive
galaxies are more likely to have B/P bulges inside their bars. Only
$\sim 12$\% of barred galaxies with stellar masses $\Mstar < 10^{10.4}
\Msun$ have B/P bulges, while $\sim 80$\% of those with higher masses
do. Correlations also exist between B/P bulge frequency and Hubble type,
gas mass ratio, galaxy colour, and bar size, in the sense that B/P hosts
are earlier in Hubble type, lower in gas fraction, redder, and have
larger bars. However, these are mostly if not entirely side effects of
the underlying correlations of these parameters with stellar mass; they
largely vanish if we control for stellar mass. A logistic fit to the B/P
bulge fraction as a function of stellar mass suggests a transition mass
of $\logmstar = 10.37$, at which point 50\% of barred galaxies have B/P
bulges.

We tested the validity of our B/P bulge--stellar mass correlation by
combining the dataset of \citet{yoshino15}, who identified B/P bulges in
a large sample of SDSS images of edge-on galaxies, with stellar-mass
estimates from the MPA-JHU DR7 database. This showed the same trend: the
B/P bulge fraction (in all edge-on galaxies) increases with stellar
mass, even after accounting for resolution effects. We also attempted to
determine the B/P bulge fraction in \textit{barred} galaxies for their
sample by using their analysis of face-on galaxies to estimate the bar
fraction as a function of stellar mass, and then applying this to the
edge-on galaxies. The result was a broadly similar trend, with evidence
that the B/P bulge fraction in barred galaxies reaches 50\% at
$\logmstar \approx 10.3$--10.4, very similar to the value from our
sample.

Although a number of simulations have suggested that gas mass fractions
$\ga 0.05$ in galaxy discs can suppress the vertical buckling of bars
(which gives rise to B/P bulges), we find no evidence for a correlation
between gas mass ratio (the ratio \fgas{} of neutral
atomic gas to stellar mass) and the presence or absence of B/P bulges,
once the correlation with stellar mass is accounted for. B/P bulges are
present in barred galaxies with frequencies as high as $\sim 80$\% for
\fgas{} as large as 0.1. This may be an indication that many B/P bulges
form via \textit{symmetric} growth mechanisms (which are less affected
by high \fgas), rather than by the buckling instability.

The sizes of B/P bulges (half-length measured along the major axis) can
be described in terms of linear size -- ranging from 450 pc to 4.6 kpc, with a
mean of $2.1 \pm 1.0$ kpc -- and as a fraction of the whole bar's length.
Since bar lengths are somewhat uncertain and difficult to define, we
measured two lengths: \amax{} and \lbar. The relative sizes
$\rbox/\amax$ and $\rbox/\lbar$ are $\rbox/\amax = 0.53 \pm 0.12$ and
$\rbox/\lbar = 0.42 \pm 0.09$, respectively. These are consistent with
previous measurements. We find no real evidence for correlations between
relative B/P bulge size and host galaxy properties.

Finally, we note that given the Milky Way's stellar mass and the
presence of a bar, the existence of a B/P bulge is entirely expected
($\sim 85$\% of local barred galaxies with similar masses have B/P
bulges), though the relative size of Galaxy's B/P bulge is very near
the lower end of observed $\rbox/\lbar$ values.

\section*{Acknowledgments} 

We are happy to thank Preben Grosb{\o}l for providing $K$-band images of
galaxies from \citet{grosbol04}. We also thank Dave Wilman for helpful
comments and advice, and Jerry Sellwood and Fran\c{c}oise Combes for
comments on earlier versions; we also thank the anonymous referee for
several suggestions that improved the readability and flow of the paper.
P.E. apologizes to V.P.D. for removing the word ``hw{\ae}t'' from our
previous paper. V.P.D. was supported by STFC Consolidated grant
\#~ST/M000877/1. VPD acknowledges being a part of the network supported
by the COST Action TD1403 ``Big Data Era in Sky and Earth Observation.''

This research is based on observations made with the NASA/ESA
\textit{Hubble Space Telescope}, obtained from the data archive at the
Space Telescope Institute.  STScI is operated by the association of
Universities for Research in Astronomy, Inc.\ under the NASA contract
NAS 5-26555.  This research also made use of both the NASA/IPAC
Extragalactic Database (NED) which is operated by the Jet Propulsion
Laboratory, California Institute of Technology, under contract with
the National Aeronautics and Space Administration, and the Lyon-Meudon
Extragalactic Database (LEDA; http://leda.univ-lyon1.fr).

This work is based in part on observations made with the
\textit{Spitzer} Space Telescope, obtained from the NASA/IPAC Infrared
Science Archive, both of which are operated by the Jet Propulsion
Laboratory, California Institute of Technology under a contract with the
National Aeronautics and Space Administration. This paper also makes use of
data obtained from the Isaac Newton Group Archive which is maintained as
part of the CASU Astronomical Data Centre at the Institute of Astronomy,
Cambridge.

Funding for the creation and distribution of the SDSS
Archive has been provided by the Alfred P. Sloan Foundation, the
Participating Institutions, the National Aeronautics and Space
Administration, the National Science Foundation, the U.S. Department of
Energy, the Japanese Monbukagakusho, and the Max Planck Society. The
SDSS Web site is \texttt{http://www.sdss.org/}.

The SDSS is managed by the Astrophysical Research Consortium (ARC) for
the Participating Institutions.  The Participating Institutions are
The University of Chicago, Fermilab, the Institute for Advanced Study,
the Japan Participation Group, The Johns Hopkins University, the
Korean Scientist Group, Los Alamos National Laboratory, the
Max-Planck-Institute for Astronomy (MPIA), the Max-Planck-Institute
for Astrophysics (MPA), New Mexico State University, University of
Pittsburgh, University of Portsmouth, Princeton University, the United
States Naval Observatory, and the University of Washington.

\bibliographystyle{mnras}

\appendix

\section{Choosing the Best Method for Finding B/P Bulges}\label{app:methods}

Four basic methods have been used in the literature to find B/P bulges
in disc galaxies. In this section we discuss the advantages and
disadvantages of each in order to motivate our use of the ``box+spurs
morphology'' approach (fourth in the list).

\begin{enumerate}

\item \textbf{Edge-on detection of B/P bulges}
\citep[e.g.,][]{jarvis86,shaw87,de-souza87,dettmar90,lutticke00a,yoshino15}:
Inspection of images of edge-on galaxies is, of course, how B/P bulges
were first identified, and this method has the advantage of being direct
-- B/P bulges \textit{do} protrude out of the disc plane and are thus
identifiable when seen at the right orientation -- but is
fundamentally limited by a number of problems.

The most serious problem, for our purposes, is that it can only be used
for \textit{positive} identifications. Observing a boxy or peanut-shaped
bulge in an edge-on galaxy tells us that there is a B/P bulge present,
but observing an \textit{elliptical} bulge may mean we are seeing a B/P
bulge in a bar oriented nearly end-on, or just a classical bulge, or
even both at once.

The second serious problem with using edge-on galaxies is the difficulty
in finding and measuring \textit{in-plane} bars. This is important for
two reasons. First, we want to know what fraction of bars \textit{lack}
B/P bulges, but this requires being able to identify planar bars in
edge-on galaxies whether or not they have B/P bulges. The second reason
is that in galaxies that \textit{do} have B/P bulges, we want to be able
to measure the in-plane bar \textit{sizes}, so we can estimate what
fraction of the bar is vertically thickened. While it is at least
sometimes possible to identify in-plane bars and measure their sizes
for edge-on galaxies using the edges of plateaus and local bumps in
surface-brightness profiles
\citep[e.g.,][]{wakamatsu84,hamabe89,lutticke00b,kormendy04,bureau05}, it remains
unclear how to separate edge-on planar bars from edge-on lenses
surrounding bars (or lenses without bars at all) or from edge-on spirals
and rings (see, for example, the discussions of NGC~4762 in
\citealt{wakamatsu84} and \citealt{athanassoula05}).

The bar-orientation problem also makes it more difficult to determine
the sizes of B/P bulges themselves, since a boxier and less
peanut-shaped structure seen side-on may not be clearly distinguishable
from a more peanut-shaped bulge seen at an intermediate angle. Put
simply, since we cannot accurately determine the orientations of bars in
edge-on galaxies, we can only measure lower limits on B/P sizes.

We note that it \textit{is} possible to identify the \textit{presence}
of an in-plane bar with gas or stellar kinematics. However,
gas-kinematic diagnostics \citep{kuijken95,bureau99b,athanassoula99} are
only valid if there is sufficient coplanar, corotating gas within the
(potential) bar region. Stellar-kinematic diagnostics
\citep{chung-bureau04,bureau05,iannuzzi15}, on the other hand, are more
widely applicable, but they require more expensive allocations of telescope
time. It is also not clear how well one can constrain the size and
orientation of the bar from such approaches \citep[as noted
by][]{iannuzzi15}, or how to relate these measurements to the usual
bar-size measurements made for more face-on galaxies.

\item \textbf{Face-on stellar kinematics}
\citep{debattista05,mendez-abreu08,iannuzzi15}: This method has the
advantage of working for face-on galaxies, where identifying and
measuring bars (with or without B/P bulges) is simple and
straightforward. The primary disadvantage is that it requires expensive
spectroscopic observations \citep[e.g., several hours on an 8m-class
telescope per galaxy;][]{mendez-abreu08}. A lesser disadvantage is that
it works best for galaxies close to face-on (i.e., inclinations $\la
30\degr$), which limits the number of galaxies that can be analyzed.

\item \textbf{``Barlens'' identification}
\citep{laurikainen14,athanassoula15}: This technique relies on the
identification of barlenses -- rounder, ``lens-like'' components inside
bars \citep{laurikainen11,laurikainen13,herrera-endoqui15} -- as the
face-on or moderately inclined projections of B/P bulges. It is in
principle applicable to both face-on and moderately inclined galaxies,
allowing bar detection and measurement and also overcoming the
limited-sample-size disadvantages of the face-on stellar-kinematics
technique. Although this is a promising approach, there are some
uncertainties. First, it is not clear to us \textit{how} to reliably
identify barlenses, in part because the definition of ``lens'' is
somewhat ambiguous.\footnote{Classically, lenses have been defined as
shallow or even flat brightness profiles with steeper declines outside
\citep[e.g.,][]{kormendy79a}. But many barlenses have steep
quasi-exponential profiles with shallower profiles outside
\citep[e.g.,][]{laurikainen14}.} Second, it is not clear if barlens-like
morphologies are \textit{always} due to B/P structure. Both spheroidal
classical bulges and disky pseudobulges can be found inside bars
\citep[e.g.,][]{erwin15a}; these could potentially be identified as
barlenses. For example, some of the barred-galaxy models with classical
bulges in \citet{saha13} show very similar face-on morphologies both
before and after the B/P bulge has formed (e.g., their model RHG097). We
also note the case of NGC~3945: this barred S0 was classified as having
barlens by \citet{laurikainen11}, but that feature is clearly due to a
large inner or nuclear disc, with an embedded nuclear ring and secondary
bar \citep[e.g.,][]{erwin99,erwin03-id,cole14,erwin15a}.\footnote{We suspect that
NGC~3945, as a massive barred galaxy, probably \textit{does} have a B/P
bulge, but distinguishing it from the projected outer bar and
inner/nuclear disc would be difficult given the bar's orientation along
the galaxy minor axis.} 

\item \textbf{Box+spurs morphology in moderately inclined galaxies}
\citep{erwin-debattista13}: Finally, there is our preferred method,
which is the examination of isophote shapes in moderately inclined galaxies where the
bar is oriented away from the galaxy minor axis, looking for a
particular pattern we refer to as ``box+spurs''. 

In \citet{erwin-debattista13}, we showed, using $N$-body simulations of
barred galaxies, that the projections at intermediate inclinations of a
vertically thin, unbuckled bar and of a bar with a B/P bulge differ in
recognizable ways (Figure~\ref{fig:box-spurs-demo}). The projection of a
thin bar produces fully symmetric, elliptical isophotes, while a bar
with a B/P bulge produces thicker, elliptical or boxy isophotes (the
``box'', which is the B/P bulge itself) and thinner, usually
\textit{offset} isophotes outside (the ``spurs'', due to the vertically
thin outer part of the bar). This is an example of a general phenomenon
noted by, e.g., \citet{stark77}: projections of concentric, aligned 3D
ellipsoids with varying axis ratios produce isophotes with varying
\textit{position angles} (isophote twists). Since the vertical axis
ratio of the B/P bulge is different from that of the outer bar, any
projection in which the galaxy is not oriented face-on (or edge-on) and
where the bar is not precisely along either the major or minor axis of
the projected disc will produce twisted isophotes inside the bar. As
demonstrated in \citet{erwin-debattista13}, this twist has a
characteristic pattern: the inner, fatter isophotes of the
box\footnote{The isophotes of the box are often quite rectangular, but
can also be more elliptical.} (from the projection of the B/P bulge) are
closer to the disc major axis than the outer, more elongated isophotes
of the spurs (from the projection of the outer bar). But since
vertically thin bars have, to first order, axis ratios which do not
change with radius, their projected isophotes do not show systematic
twists (see bottom row of panels in Figure~\ref{fig:box-spurs-demo}).

\citet{erwin-debattista13} found that projected B/P bulges in $N$-body
models could be detected for inclinations $\ga 40\degr$ and
relative angles between the bar and the disc major axis of $\la
70\degr$. There is no practical upper limit on the inclination --
indeed, we have known for some time that we can detect B/P bulges in
edge-on galaxies! -- except that for inclinations $\ga 70\degr$ it
becomes harder to reliably identify and measure the bar as a whole, and
then we are back to the problems of the edge-on approach.

The fundamental advantage of using moderately inclined galaxies is that
the box+spurs morphology is a direct consequence of the different
vertical thicknesses of the B/P structure and the outer bar. That is, when
we see the box+spurs morphology, we are assured that the ``box'' region is
vertically thicker than the spurs, and that we have a genuine B/P structure
within the bar. (The fundamental disadvantage is that this only works if the
bar is not too close to the minor axis, and so this cannot be used for \textit{all}
moderately inclined barred galaxies.)

\end{enumerate}

\section{Data Sources and Derived Quantities}\label{app:data}

For the majority of our galaxies, we used \textit{Spitzer} IRAC1
(3.6$\mu$m) images, primarily from the Spitzer Survey of Stellar
Structure in Galaxies \citep[\sfourg; ][]{sheth10} but also from the
Spitzer Infrared Nearby Galaxies Survey \citep[SINGS;][]{kennicutt03}
and other surveys \citep[e.g.,][]{dale09}. We preferred the 3.6$\mu$m
channel over the 4.5$\mu$m channel because it offers slightly higher
spatial resolution. When they were available, we also used
higher-resolution $K$-band images from \citet{knapen03},
\citet{grosbol04} and the Near-infrared Atlas of S0-Sa Galaxies
\citep[NIRS0S;][]{laurikainen11}, as well as \textit{Hubble Space
Telescope} NICMOS2 and NICMOS3 near-IR images (usually in the F160W
filter) from the \textit{HST} archive.

Accurate galaxy distances are important for absolute size measurements
and for determining stellar masses. We preferred using direct
distance measurements -- typically surface-brightness-fluctuation,
Cepheid, or TRGB measurements -- wherever possible, falling back first
to Tully-Fisher estimates when direct distances did not exist, and then
using Hubble-flow distances when T-F distances were not possible.
Sources for individual galaxies are listed in
Table~\ref{tab:parent}. For Tully-Fisher distances, we have chosen to
standardize on the 3.6\micron{} relation and distances of
\citet{sorce14}, which we use for 56 of our galaxies. Distances for
other galaxies without primary measurements but which had 3.6\micron{}
magnitudes and \hi{} velocity widths were estimated using Eqn.~1 of
\citet{sorce14}. Finally, for the remaining galaxies we determined
distances using the HyperLeda redshifts (corrected for Virgo-centric
infall) and a Hubble constant of  $\ho = 72$ km~s$^{-1}$~Mpc$^{-1}$.

For most of our galaxies, we determined the stellar mass using 2MASS
total $K$-band absolute magnitudes and stellar $M/L$ ratios estimated
from HyperLeda total corrected \bmv{} colours via the relations in
\citet{bell03}. Total \bmv{} colours were not
available for some galaxies, so we use HyperLeda \bmvre{} colours (measured within the
effective radius) instead. Six galaxies had no \bmv{} values in
HyperLeda at all, but \textit{did} have SDSS \gmr{} colours, which we
converted to \bmv{} using the Lupton (2005) 
relations.\footnote{\url{http://classic.sdss.org/dr7/algorithms/sdssUBVRITransform.html}}
Finally, there were a total of ten galaxies for which suitable optical
colours (or, in the case of two galaxies, total $K$-band magnitudes)
could not be found; for these, we used stellar-mass estimates based on
\textit{Spitzer} 3.6\micron{} and 4.5\micron{} data from
\citet{zaritsky14} or \citet{cook14}, after correcting for any
differences in the assumed distances. (We investigated using the
Zaritsky et al.\ masses for more galaxies, since they would provide mass
estimates independent of the $K$-band magnitudes and optical-colour-based
$M/L$ ratios, but found that only 29 out of 84 galaxies in our GoodPA
subsample galaxies were in that study.) All colours and magnitudes were
corrected for Galactic extinction using the extinction coefficients of
\citet{schlafly11}, as tabulated in NED.

Because the gas mass fraction of the galaxy disc has been shown to
weaken or suppress buckling in some simulations
\citep{berentzen98,debattista06,berentzen07,wozniak09}, we also gathered
data on the abundance of gas in our galaxies. For this we used the
HyperLeda corrected \hi{} magnitudes \mhi. We converted this to gas mass
in solar masses using the following relation from \cite{giovanelli88}: 
\begin{equation} 
\MHI = 2.356 \times 10^{5} D^{2} 10^{0.4 (17.40 - \mhi)}, 
\end{equation} 
where $D$ is the distance in Mpc. We found HyperLeda \hi{} measurements
for 168 of the 186 galaxies in our parent sample. To supplement this, we
searched for \hi{} data in NED and the literature for those galaxies
lacking this data in HyperLeda; we found one additional detection
\citep[NGC~3945;][]{serra12} and 16 additional upper limits, and one
galaxy (NGC~1537) for which we could find no \hi{} observations at all.
We multiply the \hi{} masses by 1.4 to account for presence of He; we
defined the gas mass fraction \fgas{} to be the ratio of atomic (H + He)
gas to stellar mass. Ideally, we should also include molecular gas, but
CO measurements are generally still too rare to be useful for a sample
of our size, so we limited ourselves to atomic gas masses.

For most of our analysis, we treated the upper limits on \hi{} as actual
measurements; however, we also computed statistical tests for alternate subsamples
that exclude the upper-limit galaxies entirely.

As noted above, we have \bmv{} colours for almost all our sample. We also
computed \bmk{} colours by combining HyperLeda $B_{tc}$ and 2MASS
$K_{s,{\rm tot}}$ magnitudes.

For barred galaxies, we measured bar position angles and sizes using a
combination of ellipse fits to isophotes (using the standard
\texttt{ellipse} task in \textsc{iraf}) and direct visual inspection. We
recorded two radial bar-size measurements based on the approach of
\citet{erwin05b}. The first is \amax, the semi-major axis where the
fitted ellipses reach maximum ellipticity -- or, in some cases, the
semi-major axis where the position angle reaches a local extremum. The
second is \lbar, which is the minimum of three measurements: \aten, the
semi-major axis where fitted ellipses deviate from the bar's position
angle by $\ge 10\degr$; \amin, the local minimum in ellipticity outside
\amax; and the distance from the galaxy centre to spiral arms or rings
which visibly cross the ends of the bar. These two measurements
function as lower and upper limits on the bar size.

\section{New B/P Galaxies}\label{app:new-bp-plots} 

In Figure~\ref{fig:new-bp-plots} we show isophote plots for 27 galaxies
with newly identified B/P bulges, along with six galaxies which were
mentioned as B/P hosts in \citet{erwin-debattista13} but for which no
figures were presented (two of these -- NGC~4442 and NGC~7582 -- were
originally identified as B/P hosts by \citealt{bettoni94} and
\citealt{quillen97}). The plots show logarithmically scaled near-IR
images, with red arrows indicating the position angle and linear extent
($2 \times \rbox$) of the B/P bulge. Cyan arrows indicate the position
angle of the outer part of the bar; the inner arrowheads indicate our
lower-limit estimate of the bar size (\amax), while the outer, more
elongated arrowheads indicate our upper-limit estimate (\lbar). Finally,
dashed black lines show the line of nodes for the galaxy disc.

Unless otherwise specified, data are \textit{Spitzer} IRAC1
(3.6\micron) images from \sfourg; in the majority of cases, images have
been smoothed using median filters with widths of 3--9 pixels. Sources
for non-\sfourg{} data: NGC~925 and NGC~1097: IRAC1 images from
\citealt{kennicutt03}; NGC~1350, NGC~1512, NGC~1537 and NGC~2781:
$K$-band images from \citealt{laurikainen11}; NGC~1371: $K$-band image
from \citealt{grosbol04}; NGC~2903: IRAC1 image from \citealt{dale09};
NGC~6744: IRAC1 image from \textit{Spitzer} archive [PI David Fisher,
Program ID 30496]; NGC~7582: $H$-band image from \citealt{eskridge02}.

\begin{figure*}
\begin{center}
\vspace*{-3mm}
\includegraphics[scale=0.9]{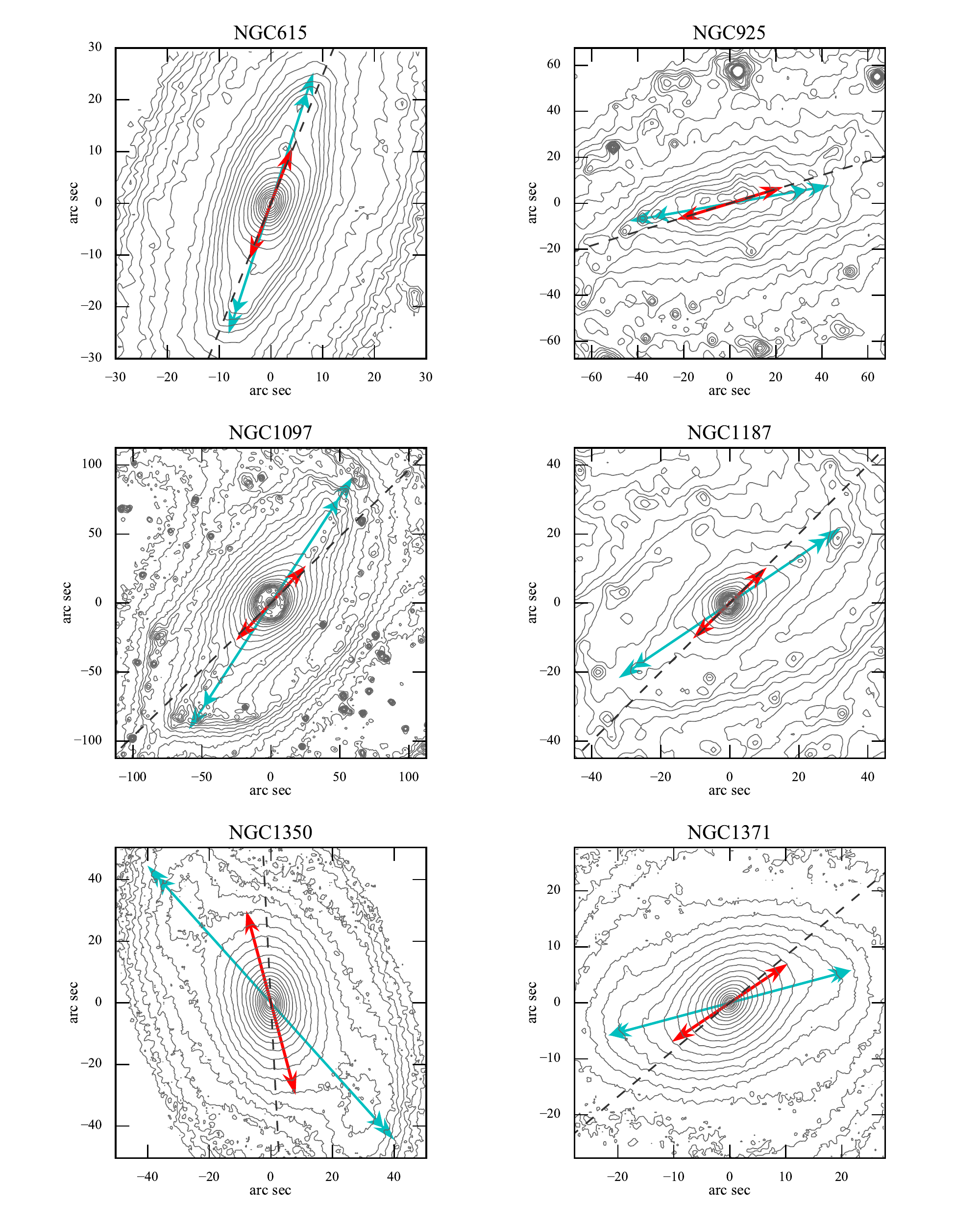}
\end{center}

\caption{Log-scaled near-IR isophote contours of the bar regions in
galaxies with newly-identified B/P bulges or galaxies listed in
\citet{erwin-debattista13} with B/P bulges but no figures. Red arrows
indicate PA and size of projected B/P bulge (length = $2 \times \rbox$);
cyan double-headed arrows indicate PA and lower and upper limits on size
of bar (inner arrow: $2 \times \amax$; outer arrow: $2 \times \lbar$).
Dashed black line indicates disc major axis.\label{fig:new-bp-plots}}

\end{figure*}

\addtocounter{figure}{-1}
\begin{figure*}
\begin{center}
\vspace*{-1cm}\includegraphics[scale=0.9]{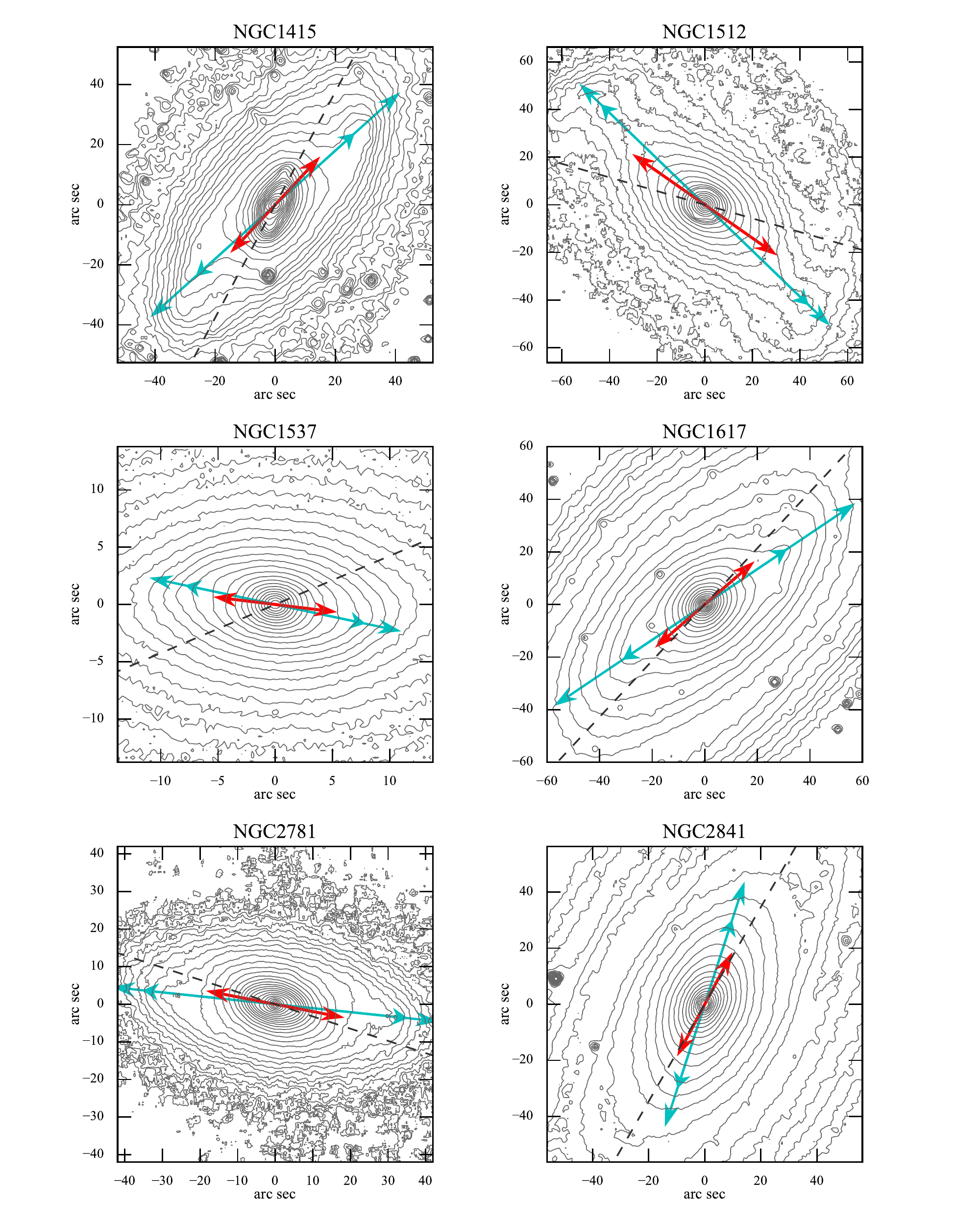}
\end{center}

\caption{-- continued.}

\end{figure*}

\addtocounter{figure}{-1}
\begin{figure*}
\begin{center}
\vspace*{-1cm}\includegraphics[scale=0.9]{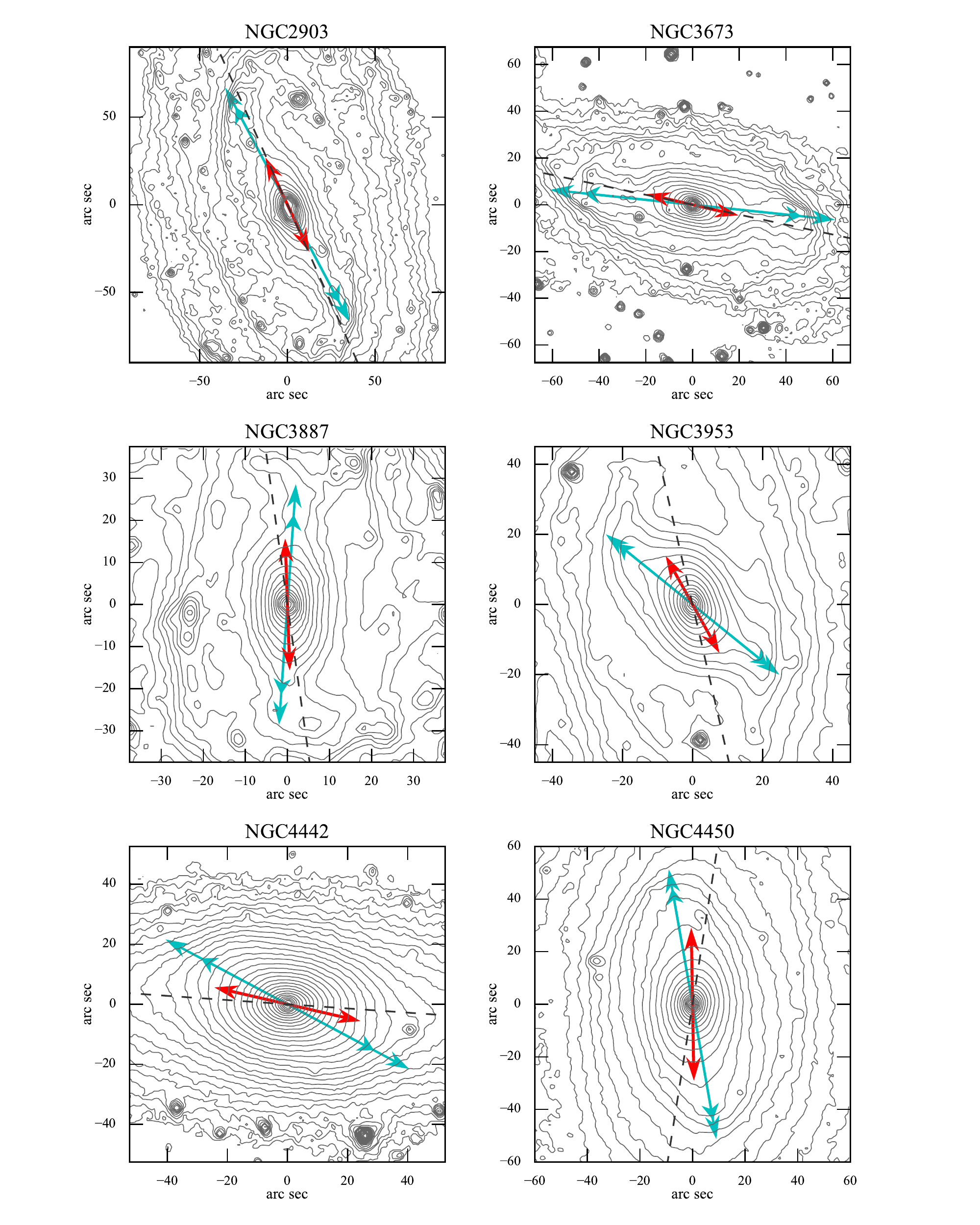}
\end{center}

\caption{-- continued.}

\end{figure*}

\addtocounter{figure}{-1}
\begin{figure*}
\begin{center}
\vspace*{-1cm}\includegraphics[scale=0.9]{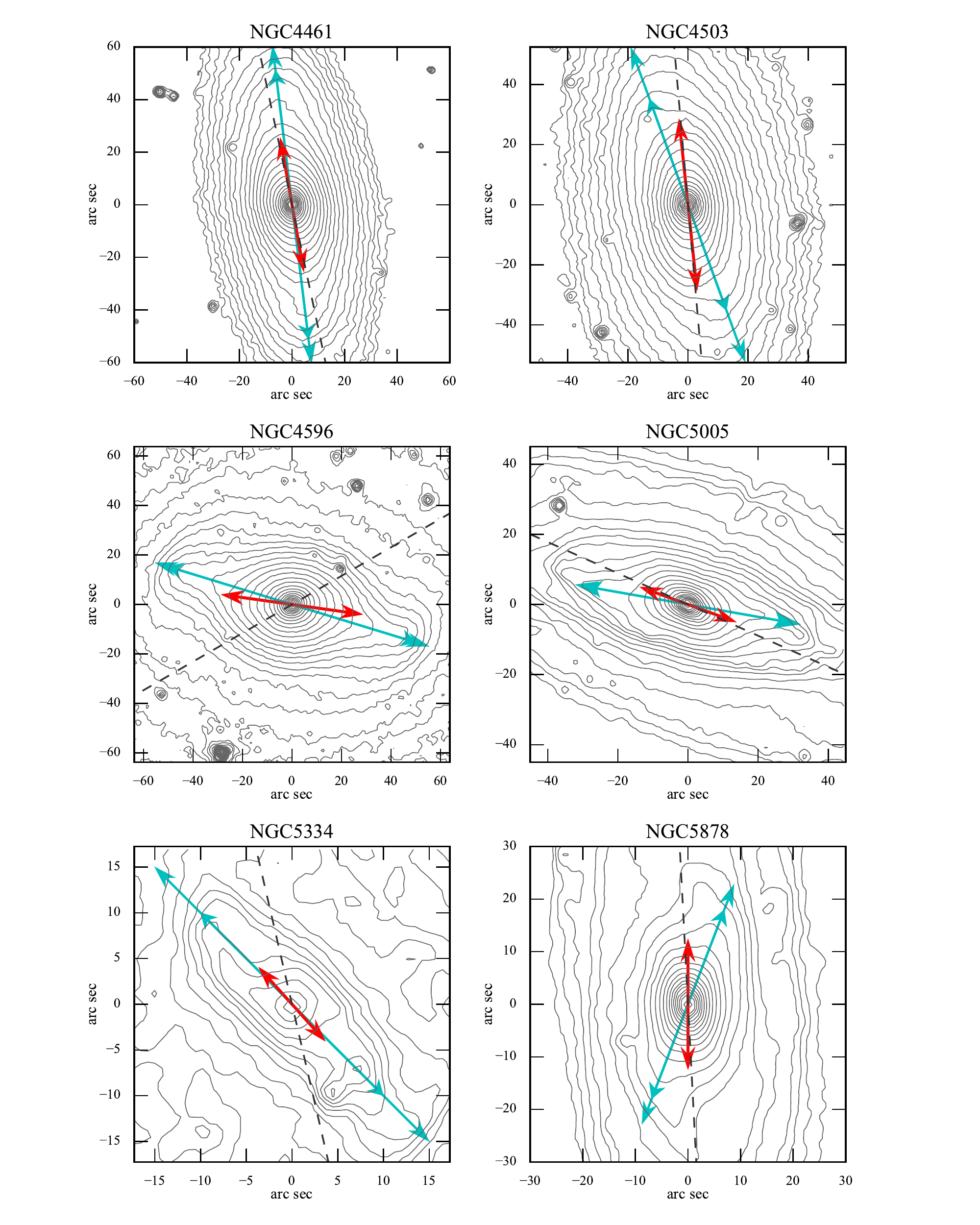}
\end{center}

\caption{-- continued.}

\end{figure*}

\addtocounter{figure}{-1}
\begin{figure*}
\begin{center}
\vspace*{-1cm}\includegraphics[scale=0.9]{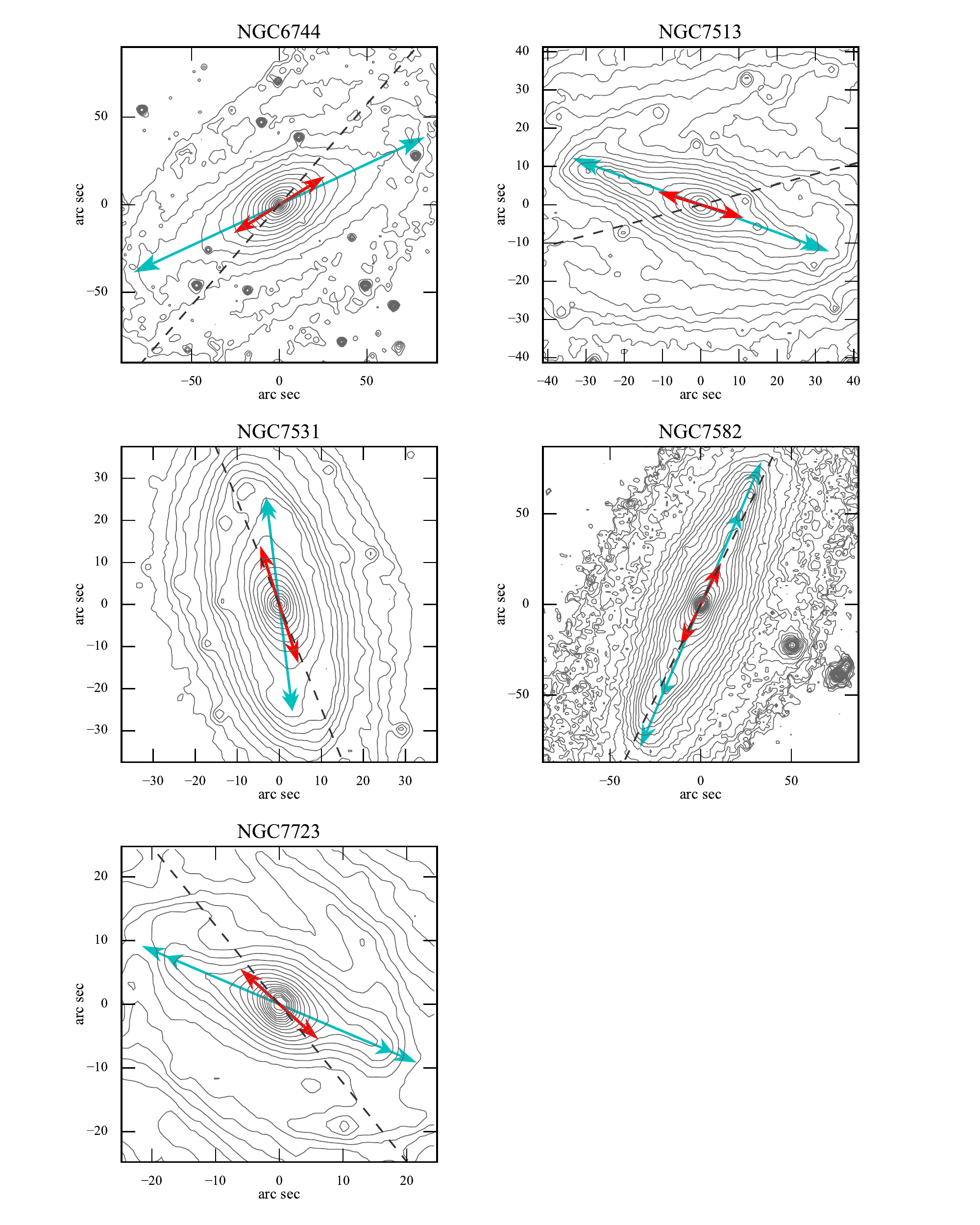}
\end{center}

\caption{-- continued.}

\end{figure*}

\section{Parent Sample}\label{app:parent-sample} 

\begin{table*}
\begin{minipage}{126mm}
    \caption{Parent Sample}
    \label{tab:parent}
    \begin{tabular}{lllrrrrrrrrr}
\hline
Name       & Type (RC3) & Bar & Distance & Source & $i$  & PA   & $M_{K}$ & $\log \Mstar$ & Source & $\log \MHI$ & Source \\
           &            &     & (Mpc)    &        & (\degr) & (\degr) &   & (\Msun)       &        & (\Msun)     & \\
(1)        & (2)        & (3) & (4)      & (5)    & (6)  & (7)  & (8)     & (9)           & (10)   & (11)        & (12) \\
\hline
NGC150         & SB(rs)b    &  Y & 19.70    & 1     &  62 & 117 & $-22.97$ & 10.39 & 1 &  9.43 & 1 \\
NGC157         & SAB(rs)bc  &  N & 12.90    & 2     &  48 &  50 & $-22.88$ & 10.34 & 1 &  9.25 & 1 \\
NGC210         & SAB(s)b    &  Y & 21.60    & 2     &  49 & 160 & $-23.29$ & 10.53 & 1 &  9.77 & 1 \\
NGC300         & SA(s)d     &  N &  2.00    & 3     &  50 &  20 & $-20.13$ &  9.25 & 1 &  9.29 & 1 \\
NGC450         & SAB(s)cd   &  Y & 16.60    & 4     &  56 &  77 & $-20.53$ &  9.39 & 1 &  9.10 & 1 \\

\hline
\end{tabular}

\medskip

(1) Galaxy name. (2) Hubble type from RC3. (3) Bar present. (4) Distance in Mpc. (5) Source of distance: 
1 = HyperLeda redshift (corrected for Virgo-centric infall); 2 = Tully-Fisher from
literature data, using \citet{sorce14} Eqn.~1; 3 = Cepheids \citep{freedman01};
4 = Tully-Fisher \citep{sorce14}; 5 = SBF distance from \citet{tonry01}, including 
metallicity correction from \citet{mei05}; 6 = Cepheids \citep{qing15}; 7 = default 
Fornax Cluster distance \citep{blakeslee09};
8 = Cepheids \citep{riess16}; 9 = SBF \citep{blakeslee09}; 10 = mean of T-F
distances from \citet{theureau07}; 11 = Cepheids \citep{macri01}; 12 = TRGB \citep{dalcanton09};
13 = SBF distance of interacting neighbor NGC~3226 \citep{tonry01}; 14 = SBF \citep{mei07};
15 = default Virgo Cluster distance \citep{mei07}; 16 = TRBG (mean of NED values);
17 = Tully-Fisher \citep[][, erratum]{springob09}; 
18 = Tully-Fisher \citep[mean of][]{springob09,nasonova11,tully13}.
(6) Absolute 2MASS total $K$ magnitude, retrieved
from NED and using our adopted distance. (7) Inclination. (8) Position angle
(deg E from N) of disk major axis. (9) Stellar mass. (10) Source for stellar mass:
1 = optical color + \citet{bell03}; 2 = \citet{zaritsky14}; 3 = \citet{cook14}.
(11) Neutral gas mass. (12) Source for neutral gas mass: 1 = HyperLeda; 2 = NED;
3 = \citet{serra12}, 4 = \citet{giovanardi83,boselli14}. The full table is available in
the online version of the paper; we show a representative sample here.
   
\end{minipage}
\end{table*}

The final parent sample for our analysis (Parent; see Section~\ref{sec:sample})
is listed in Table~\ref{tab:parent}.  Galaxies which were rejected during the
construction of this sample are listed in Appendix~\ref{app:rejected}, along
with the reasons for rejection.

\section{Galaxies with Nuclear Bars But No Large-Scale Bars}\label{app:nuclear-bars}

Three early- or intermediate-type galaxies in our sample -- NGC~1201
(S0), NGC~1553 (S0), and NGC~5194 (Sbc) -- do not have large-scale bars, but
\textit{do} have very small ``nuclear'' bars. Although the term ``nuclear bar''
is poorly defined, the bars in these galaxies have
deprojected sizes $\amax \lesssim 0.05 \, R_{25}$ ($\amax/R_{25} =$ 0.047,
0.039, and 0.042 for NGC~1201, NGC~1553, and NGC~5194, respectively).
This places them below the range spanned by S0--Sb galaxy bars in
\citet{erwin05b}, and below the range of S0--Sc bars in the
study of \citet{menendez-delmestre07}, with the exception of NGC~5194
itself. The next smallest bar in the S0--Sc galaxies of
Men{\'e}ndez-Delmestre et al\ is in NGC~1068, with $\amax/R_{25} =
0.07$, which is in fact the \textit{inner} bar of a double-bar system
\citep[][and references therein]{erwin04}. On this basis, we classify
these three galaxies as ``unbarred'' for the purposes of our study.

\section{Rejected Galaxies}\label{app:rejected} 

The following galaxies met our initial RC3-based criteria
(Section~\ref{sec:sample}), but were excluded from the final Parent
Sample. In the majority of cases, we determined that the galaxies had
inclinations lower or higher than our limits ($i = 40\degr$--70\degr), in
spite of having 25th-magnitude axis ratios in RC3 which suggested
otherwise. These include cases of edge-on galaxies with luminous bulges
or halos (leading to low RC3 axis ratios) as well as near-face-on
galaxies with large, luminous bars, lenses, and/or elliptical inner
rings, where the RC3 axis ratio is due to the bar/lens/ring.

\textbf{NGC~289:} Ellipse fits to the outer disc (Sptizer IRAC1 image)
indicate an inclination of 38\degr.

\textbf{NGC~613:} Various kinematic analyses in the literature suggest
an inclination of $\sim 35\degr$.

\textbf{NGC~986:} Analysis of a \textit{Spitzer} IRAC1 image by \citet{munoz-mateos13}
found $i = 72\degr$.

\textbf{NGC~988:} Removed due to the presence of an extremely bright
star within the galaxy.

\textbf{NGC~1042:} Removed due to low inclination \citep[$i = 38\degr$
from outer-disc ellipse fits,][]{pt06}.

\textbf{NGC~1300:} The analysis of \hi{} data in \citet{lindblad97} indicates an inclination
of 35\degr.

\textbf{NGC~1316:} Although this galaxy is classified as S0 in RC3, it is a well-known merger
remnant \citep[e.g.][]{goudfrooij01a,goudfrooij01b}.

\textbf{NGC~1543:} The outer isophotes of this galaxy indicate an
inclination of $\sim 20\degr$ \citep{erwin15a}, well below our lower
limit; the RC3 axis ratio is due to the outer bar and lens.

\textbf{NGC~2146:} We rejected this galaxy due to its status as a clear
merger remnant.

\textbf{NGC~2685:} This is a well-known polar-ring galaxy.

\textbf{NGC~2775:} Removed due to low inclination \citep[$i = 39\degr$ from outer-disc 
ellipse fits,][]{gutierrez11}.

\textbf{NGC~2805:} The analysis of Fabry-Perot \ha{} data in
\citet{epinat08} indicates an inclination of $\approx 17\degr$.

\textbf{NGC~3027:} The analysis of Fabry-Perot \ha{} data in
\citet{epinat08} indicates an inclination of $\approx 77\degr$.

\textbf{NGC~3310:} Removed for being a probable merger remnant
\citep[e.g.,][]{kregel01,wehner05}.

\textbf{NGC~3403:} Ellipse fits to the \sfourg{} 3.6\micron{} image indicate
an inclination of $i \approx 71\degr$, slightly outside our limits.

\textbf{NGC~3414:} Ellipse fits to the \sfourg{} 3.6\micron{} image indicate an
inclination of $\approx 34\degr$.

\textbf{NGC~3607:} Analysis of $R$-band isophotes indicates an inclination of
$\approx 29\degr$ \citep{gutierrez11}.

\textbf{NGC~3630:} The morphology of this galaxy clearly suggests an edge-on S0, 
despite the RC3 axis ratio.

\textbf{NGC~3718:} This is a complex galaxy almost certainly the result of a
relatively recent interaction, and possibly related to polar-ring galaxies. Without
a well-defined disc, we cannot include it.

\textbf{NGC~3733:} Ellipse fits to the \sfourg{} IRAC1 image suggest an
inclination of $\approx 71\degr$.

\textbf{NGC~3755:} The analysis of Fabry-Perot \ha{} data in
\citet{epinat08} indicates an inclination of $\approx 77\degr$.
(Ellipse fits to SDSS images suggest an inclination of $\sim 74\degr$.)

\textbf{NGC~3981:} Ellipse fits to the \sfourg{} IRAC1 image suggest an
inclination of $\approx 75\degr$.

\textbf{NGC~4036:} The morphology of this galaxy clearly suggests an
edge-on S0, despite the RC3 axis ratio.

\textbf{NGC~4051:} Analysis of \hi{} data by \citet{liszt95} suggests an inclination
of 37\degr.

\textbf{NGC~4151:} The inclination of this galaxy is only $\sim 20\degr$
\citep[e.g.,][]{erwin05b}; the RC3 axis ratio is due to the large bar + lens.

\textbf{NGC~4251:} The morphology of this galaxy clearly suggests an edge-on S0, 
despite the RC3 axis ratio.

\textbf{NGC~4258:} \hi{} velocity-field analysis suggests an inclination of
72\degr{} for this galaxy \citep{van-albada80}. 

\textbf{NGC~4350:} The morphology of this galaxy clearly suggests an edge-on S0, 
despite the RC3 axis ratio.

\textbf{NGC~4382:} This galaxy lacks a clearly defined outer disc, and is probably
a merger remnant \citep[e.g.,][and references therein]{gutierrez11}.

\textbf{NGC~4417:} The morphology of this galaxy clearly suggests an edge-on S0, 
despite the RC3 axis ratio.

\textbf{NGC~4424:} Removed due to being a likely merger remnant \citep[e.g.,][]{kenney96}.

\textbf{NGC~4441:} Removed for being a clear merger remnant \citep[e.g.,][]{manthey08,jutte10}.

\textbf{NGC~4459:} Removed due to low inclination \citep[$i = 38\degr$ from outer-disc 
ellipse fits,][]{gutierrez11}.

\textbf{NGC~4488:} This galaxy has a genuinely peculiar morphology, with a box-shaped
interior, two elongated (tidal?) spiral arms, and no clear outer disc. Since we cannot
determine a reliable orientation, for this galaxy, we exclude it.

\textbf{NGC~4490:} This galaxy is strongly interacting with its neighbor NGC~4485,
making determination of its orientation too difficult.

\textbf{NGC~4539:} Ellipse fits to the \sfourg{} IRAC1 image suggest an
inclination of $\approx 75\degr$.

\textbf{NGC~4594:} This is the Sombrero Galaxy, an almost edge-on Sa with a luminous
bulge which produces the relatively round RC3 axis ratio.

\textbf{NGC~4643:} Removed due to low inclination \citep[$i = 38\degr$ from outer-disc 
ellipse fits,][]{erwin08}.

\textbf{NGC~4699:} Analysis of the outer-disc isophotes from SDSS images yields an
inclination of $\approx 37\degr$ \citep{erwin15a}.

\textbf{NGC~4731:} This galaxy consists of a strong, very narrow bar and
two strong, open spiral arms, forming an integral-sign shape. We
excluded it because we were unable to measure a reliable outer disc
orientation.

\textbf{NGC~5248:} Publicly available images of this galaxy are not deep enough
for us to reliably determine the outer disc orientation (the analysis of \citealt{jogee02a}
suggests an inclination of $\sim 40\degr$, meaning we cannot be sure it is inclined
enough to meet our inclination criteria).

\textbf{NGC~5866:} The morphology of this galaxy clearly suggests an edge-on S0, 
despite the RC3 axis ratio.

\textbf{NGC~6255:} Ellipse fits to SDSS images indicate an inclination
of $\approx 71\degr$.

\textbf{NGC~7041:} The morphology of this galaxy clearly suggests an edge-on S0, 
despite the RC3 axis ratio.

\textbf{NGC~7412:} Ellipse fits to the \sfourg{} IRAC1 image of this galaxy suggest
an inclination of $\approx 37\degr$.

\textbf{NGC~7727:} Rejected for being a clear merger remnant.

\textbf{NGC~7814:} This is an edge-on early-type spiral with a large bulge, similar to 
the Sombrero Galaxy (NGC~4594).

\textbf{IC~4212:} No useful, publicly available imaging data for this galaxy exists.

\textbf{UGC~6930:} The optical and \hi{} analysis of this galaxy in \citet{verheijen01}
indicate an inclination of $\approx 31\degr$.

\textbf{ESO~499-37:} We were unable to determine a plausible stellar mass for this galaxy.
The combination of the LEDA $B_{tc}$ value and the available 2MASS photometry yields $\bmk = -0.5$,
suggesting the $K$ magnitude is much too faint.

\section{Notes on Individual Galaxies}\label{app:individual-galaxies} 

\textbf{NGC~2273:} We were unable to find IRAC1 total-magnitude measurements for
this galaxy, so we used the mean of the $JHK$ Tully-Fisher measurements from \citet{theureau07}
for the distance.

\textbf{NGC~3227:} We use the SBF distance of this galaxy's interacting
neighbor NGC~3226 \citep{tonry01}.

\textbf{NGC~4596:} This was classified as \textit{not} having a B/P
bulge in \citet{erwin-debattista13}. Based on the slight offset of the
spurs relative to the rounder interior of the bar (a very oval ``box''
region), we now count this as having a B/P bulge. This is consistent
with the identification of a barlens in this galaxy by
\citet{laurikainen11}.

\textbf{NGC~4772:} Attempts to use the \citet{sorce14} Tully-Fisher
relation for this galaxy yield distances of $\sim 50$--70 Mpc, depending
on the inclination; published Tully-Fisher distances in NED range from 25 to 41 Mpc.
Since the Virgocentric-infall-corrected recession velocity is only 1083 \kms{} and since
most studies tend to associate it with the Virgo Cluster, we use our default
Virgo Cluster distance (16.5 Mpc) for this galaxy.

\section{Galaxy Simulations}\label{sec:sims}

In Figure~\ref{fig:box-spurs-demo} we show projections of three
different galaxy simulations at different orientations as a way of
demonstrating the effects of having (or not having) a B/P bulge on the
observed isodensity or isophote contours of a barred galaxy. Since two
of these simulations were previously used in \citet{erwin-debattista13}
and \citet{erwin-debattista16}, we continue the naming scheme used in
those papers. Simulations~A and B are pure $N$-body simulations which
were previously discussed in \citet{erwin-debattista13}; they use
300,000--500,000 stellar disc particles with softening lengths of 60 pc
(Simulation~A) or 0.05 natural units (Simulation~B). (More details of
Simulation~B can be found in \citealt{sellwood09}.) Both simulations
formed bars which subsequently buckled; we show Simulation~A at a time
after the buckling of the bar and the formation of the B/P bulge, while
Simulation~B is shown after the bar has formed but \textit{before}
buckling, so the bar is still vertically thin.

Simulation~E\footnote{Simulations~C and D do not in appear in this
paper, but were used in \citet{erwin-debattista16}.} is an
$N$-body$+$SPH simulation first presented in \citet{ness14} and
\citet{cole14}; a complete description can be found in
\citet{debattista17}. The stars in this simulation form entirely out of
gas cooling from a spherical corona, triggering continuous star
formation.  This simulation, which was evolved with \textsc{Gasoline}
\citep{wadsley04}, the smooth particle hydrodynamics version of
\textsc{Pkdgrav}, has high force resolution (50~pc) and stellar mass
resolution ($9.5 \times 10^{3} \Msun$).  The model forms a strong bar
between 2 Gyr and 4 Gyr \citep{cole14}, while a B/P  bulge forms by
10~Gyr.

As pointed out by the referee, the \textit{unbuckled} bar in
Simulation~B (bottom left and bottom middle panels in
Figure~\ref{fig:box-spurs-demo}) shows some pinching of the isophotes
near the center of the bar. This is a feature that shows up in at least
some of our simulations of bars prior to the buckling phase. We suspect
this is a side effect of initial conditions involving overly cold disks
in the simulations, with bar formation causing in-plane contraction in
all directions, but more strongly along the minor axis; in a real
galaxy, this would probably be weakened by more efficient vertical
heating of the nuclear region of the disk. (We see no evidence of such
pinching in the observations, although this would be difficult to see
for two reasons. First, galaxies in our sample have inclinations of
40--70\degr, while the pinching in the simulations is best seen for
face-on orientations. Second, many of the observed unbuckled bars have
regions of star formation within the bars; the excess light has the
effect of making the combined, observed isophotes rounder in the bar
region, and could thus wash out any weak pinching in the bar.)

\bsp	
\label{lastpage}
\end{document}